\documentclass[12pt]{article}
\usepackage{amsmath}
\usepackage{graphicx}
\usepackage{enumerate}
\usepackage{natbib}
\usepackage{url}

\usepackage{bm}
\usepackage{booktabs}
\usepackage{amsmath}
\usepackage{graphicx}
\usepackage[colorlinks=true, allcolors=blue]{hyperref}
\usepackage[labelfont=bf, font=singlespacing]{caption}
\usepackage{multirow}
\usepackage{multicol}
\usepackage{adjustbox}
\usepackage[english]{babel}
\usepackage{amssymb}
\usepackage{amsthm}
\usepackage[table,xcdraw]{xcolor}
\usepackage{setspace}
\theoremstyle{definition}
\newtheorem{definition}{Definition}[section]
\DeclareMathOperator*{\argmax}{arg\,max}
\DeclareMathOperator*{\argmin}{arg\,min}
\usepackage{subfiles}
\usepackage{colortbl}
\usepackage[normalem]{ulem}
\useunder{\uline}{\ul}{}

\newcommand{\blind}{1}

\newcommand{\beginsupplement}{%
        \setcounter{table}{0}
        \renewcommand{\thetable}{S\arabic{table}}%
        \setcounter{figure}{0}
        \renewcommand{\thefigure}{S\arabic{figure}}%
     }

\begin{document}

\def\spacingset#1{\renewcommand{\baselinestretch}%
{#1}\small\normalsize} \spacingset{1}

\if1\blind
{
  \title{\bf Bayesian Model Averaging for Partial Ordering Continual Reassessment Methods}
  \author{Luka Kova\v{c}evi\'{c}\\
    \textit{MRC Biostatistics Unit, University of Cambridge} \medskip\\
    Thomas Jaki \\
    \textit{Department of Machine Learning and Data Science, University of Regensburg} \\ \textit{MRC Biostatistics Unit, University of Cambridge} \medskip \\
    Helen Barnett \\
    \textit{Lancaster University} \medskip\\
    and \medskip\\
    Pavel Mozgunov \\
    \textit{MRC Biostatistics Unit, University of Cambridge}}
  \maketitle
} \fi

\if0\blind
{
  \bigskip
  \bigskip
  \bigskip
  \begin{center}
    {\LARGE\bf Bayesian Model Averaging for Partial Ordering Continual Reassessment Methods}
\end{center}
  \medskip
} \fi

\bigskip
\begin{abstract}
Phase I clinical trials are essential to bringing novel therapies from chemical development to widespread use. Traditional approaches to dose-finding in Phase I trials, such as the '3+3' method and the Continual Reassessment Method (CRM), provide a principled approach for escalating across dose levels. However, these methods lack the ability to incorporate uncertainty regarding the dose-toxicity ordering as found in combination drug trials. Under this setting, dose-levels vary across multiple drugs simultaneously, leading to multiple possible dose-toxicity orderings. The Partial Ordering CRM (POCRM) extends to these settings by allowing for multiple dose-toxicity orderings. In this work, it is shown that the POCRM is vulnerable to '\textit{estimation incoherency}' whereby toxicity estimates shift in an illogical way, threatening patient safety and undermining clinician trust in dose-finding models. To this end, the Bayesian model averaged POCRM (BMA-POCRM) is proposed. BMA-POCRM uses Bayesian model averaging to take into account all possible orderings simultaneously, reducing the frequency of estimation incoherencies. The effectiveness of BMA-POCRM in drug combination settings is demonstrated through a specific instance of estimate incoherency of POCRM and simulation studies. The results highlight the improved safety, accuracy and reduced occurrence of estimate incoherency in trials applying the BMA-POCRM relative to the POCRM model. 
\end{abstract}

\noindent%
\vfill

\newpage

\setstretch{1.9}

\section{Introduction}

An aim of in-patient Phase I clinical trials is to determine the maximum tolerable dose (MTD) from a range of dose levels for progression into Phase II trials. The MTD is the dose level that matches the target probability of unwanted effects, sometimes called the target toxicity rate (TTR). Depending on the setting, the MTD can be either a dose level for a single agent or a combination of dose levels in a multi-agent setting. A common approach to determining the MTD is to use the number of dose-limiting toxicities (DLTs), simply called toxicities, on each dose to compute a probability of toxicity. The definition of a DLT and the desired TTR are set in advance of a trial, hence, by estimating the risk of DLT for each dose, the dose level with a toxicity rate closest to the TTR can be selected. Dose-escalation procedures aim to estimate the location of the true MTD given a TTR and sequential patient DLT data.

Approaches to dose-escalation in Phase I clinical trials frequently rely on the assumption of simple dose orderings. In the case of single drug Phase I trials, a simple ordering can be constructed by assuming a monotonic relationship between dose level and toxicity. That is, higher doses of a drug are assumed to be more toxic than lower doses. For example, suppose there are dose levels \(d_1, \ldots, d_4\) where a higher index indicates a higher dose level. This means that \(d_1 < \ldots < d_4\) implies the following simple ordering assuming dose-toxicity monotonicity, \begin{equation*}
d_1 \rightarrow d_2 \rightarrow d_3 \rightarrow d_4. 
\end{equation*} This also implies that \(d_2\) is more toxic than \(d_1\), \(d_3\) is more toxic than both \(d_1\) and \(d_2\) and so forth. The continual reassessment method (CRM) is one such approach that is based on the assumption of monotonicity. It utilises a Bayesian framework to update estimates of the risk of toxicity to guide dose-escalation based on a given toxicity ordering of doses \citep{o1990continual}. Several independent studies have shown that escalation based on the CRM leads to favourable operating characteristics for finding the true MTD in single agent clinical trials. With the growing need for combination drug trials, where the toxicity profile of joint administration of two or more drugs is investigated, methods that allow for potential uncertainties in the ordering of dose levels are necessary \citep{mozgunov2020surface}. 

\begin{table}[h!]
\centering
\begin{tabular}{cccc}
\hline
                        &                             & \multicolumn{2}{c}{Drug B}                                \\
                        &                             & \(1\)                     & \(2\)                     \\ \hline
\multirow{3}{*}{Drug A} & \(1\)                     & \(\bm{d_1}\)                     & \(\bm{d_2}\)                     \\
                        & \(2\)                     & \(\bm{d_3}\)                     & \(\bm{d_4}\)                     \\
                        & \multicolumn{1}{l}{\(3\)} & \multicolumn{1}{l}{\(\bm{d_5}\)} & \multicolumn{1}{l}{\(\bm{d_6}\)} \\ \hline
\end{tabular}
\caption{\(3\times2\) setting resulting in 6 dose levels.}
\label{tab:casestudy}
\end{table}

Escalation of multiple drugs concurrently creates uncertainty in the dose-escalation process as the change in dose toxicity is not obvious for diagonal transitions in dose level, for example where one drug increases in dose level and the other decreases. Specifically, it may be reasonable to assume monotonicity for a single drug, however, this does not extend to multiple changes in dose level.  In Table \ref{tab:casestudy}, the 3-by-2 dose level configuration for a dual drug combination trial with 3 dose levels for drug $A$ and 2 dose levels for drug $B$ is shown. Here, it is unknown prior to the trial whether toxicity increases or decreases with a shift between $d_{2}$ and $d_{3}$ and likewise for $d_4$ and $d_5$. Based on this dose level matrix, the 5 simple orderings comprising the partial ordering of dose levels are, \begin{align*}
    1: \; d_1 \rightarrow d_2 \rightarrow d_3 \rightarrow d_4 \rightarrow d_5 \rightarrow d_6, \\ 
    2: \; d_1 \rightarrow d_3 \rightarrow d_5 \rightarrow d_2 \rightarrow d_4 \rightarrow d_6, \\ 
    3: \; d_1 \rightarrow d_3 \rightarrow d_2 \rightarrow d_5 \rightarrow d_4 \rightarrow d_6, \\
    4: \; d_1 \rightarrow d_2 \rightarrow d_3 \rightarrow d_5 \rightarrow d_4 \rightarrow d_6, \\
    5: \; d_1 \rightarrow d_3 \rightarrow d_2 \rightarrow d_4 \rightarrow d_5 \rightarrow d_6.
\end{align*} 

Several procedures have been developed to handle the problem of uncertain dose-toxicity orderings, which is a problem that also persists in dose-schedule \citep{mozgunov2019information} and combination-schedule settings \citep{mozgunov2022practical, riviere2015competing}. \cite{ivanova2004non} developed the \emph{up-and-down design for combinations} using an isotonic regression combined with the Narayana design to escalate doses algorithmically \citep{ivanova2003improved}. A further development updated this method to utilise a \emph{T-statistic} \citep{ivanova2009dose} for escalation decisions. Further, \citet{yin2009bayesian} have developed the Bayesian copula regression-based model, which uses a Bayesian scheme similar to the CRM to update posterior estimates of toxicity and recommend dose allocation decisions. Most recently, \citet{mozgunov2020surface} proposed a beta distributed surface-free approach to handling drug combination trials, successful in reducing the average number of toxicities within a trial. The Continual Reassessment Method for Partial Orderings (POCRM), developed by \cite{wages2011continual}, extends the Bayesian framework of the CRM to allow for several dose-toxicity orderings to be specified.

Further, the POCRM addresses the problem of uncertainty in dose-toxicity orderings by selecting from the set of proposed dose-toxicity orderings for each cohort in the trial. Given a set of simple orderings that are pre-specified, POCRM selects the most likely ordering and uses this to recommend the next dose to be assigned in the trial via the CRM. This approach is particularly favourable as it allows for flexibility in the dose-toxicity ordering used for dose recommendation. However, it also significantly limits the performance of the model as, particularly in cases with a large number of dose combinations, not all possible orderings can be considered by the model since the model can only consider orderings given as candidates. It can be argued that the true MTD will be selected regardless of whether the correct dose ordering is present, however, since the proposed orderings alter toxicity estimates, they also guide dose-escalation. In particular, in the combination setting, multiple doses that have a toxicity close to the TTR may be present. Therefore, inaccurate estimation of the risk of toxicity would hide alternative true MTDs. In this case, not only is escalation important, but also point estimation. Another instance under which this method is potentially problematic is where multiple orderings have a similar posterior probability leading to the uncertainty in dose orderings being disregarded as only the model with the highest posterior probability is selected at each step. This article also explores the prevalence of illogical 'jumps' in dose-toxicity estimates present in practice, largely due to changes in the preferred simple ordering. 

To address the challenges associated with ordering selection and uncertainty quantification during trials, the original POCRM is extended by applying Bayesian model averaging (BMA) in this work. Previously, BMA has been applied to the original CRM \citep{yin2009bayesian} where it was highly successful in improving dose allocation for single-agent trials, particularly for small sample Phase I trials. The novel BMA-POCRM design aims to incorporate uncertainty in the toxicity ordering with the aim of making more flexible dose-toxicity estimates, which are not limited to following a single predefined ordering, and take into account the additional uncertainty implied by a partial ordering. 

Clinician trust is an essential aspect of running real-world adaptive clinical trials. Even if a clinical trial design has good statistical properties, as confirmed via a simulation study, the design will not be implemented in practice if its recommendations are not aligned with the evidence from the trial. A lack of trust in the design can lead to the recommendations given by the model not being followed despite existing evidence that may support the model's decision. The concept of estimation coherency, proposed and developed in this manuscript, aims to provide an additional metric for evaluating design behaviour in real-world settings. Via a case study and an extensive simulation study, it will be demonstrated that this property is crucial to ensuring that the dose-escalation and de-escalation recommendations of a design can be easily communicated to the clinical team on the trial and, hence, more likely to be adhered to. This will result in delivering more efficient early phase dose-escalation trials.

The general framework of the Bayesian CRM is outlined along with the POCRM and the novel method BMA-POCRM in Section \ref{sec:methods}. Section \ref{sec:incohstudy} provides clear motivation for the development of BMA-POCRM. Here, the novel concept of estimation incoherency is introduced, which measures the consistency of dose toxicity estimate updates with respect to the given dose toxicity orderings. Furthermore, we provide a case study of the performance of BMA-POCRM and POCRM on real trial data in Section \ref{sec:realtrial}. Together, the evidence from the case study and further simulation results in Section \ref{sec:simstudy} show that BMA-POCRM improves the accuracy, safety and operating characteristics with more intuitive escalation and de-escalation decisions. An examination of the operating characteristics is carried out in Section \ref{sec:simstudy} leading to a sensitivity analysis of results under various cohort sizes in Section \ref{sec:sensitivity}. Finally, a discussion and analysis of the results are presented in Section   \ref{sec:discussion}.

\section{Methods} \label{sec:methods}

\subsection{General Framework} \label{subsec:generalframework}

Consider a setting with a partial ordering corresponding to $M$ simple orderings and $K$ dose levels, $\{d_1, \ldots, d_K\}$. Following a framework similar to that set by the original CRM \citep{o1990continual}, let $X_j$ be the dose level assigned to the $j$-th patient where $x_j \in \{d_1, \ldots, d_K\}$ and let $Y_j$ be a binary random variable for whether patient $j$ experiences a DLT.

For a particular ordering $m\in\{1,\ldots,M\}$, the risk of DLT at $d_k \in \{d_1, \ldots, d_K\}$ is modelled as, $$  \widehat{R}(d_k) = \Pr[Y_j = 1| X_j = d_k] = \psi_m(d_k, a), $$ where $\widehat{R}(d_k)$ is the estimated risk of DLT at $d_k$, $\psi_m$ is the working model under ordering $m$ \citep{wages2011continual}, and $a$ is the model parameter. A wide range of specifications can be selected for the working model, each with their own associated assumptions regarding the dose-toxicity relationship and parameter estimation approaches \citep{cheung2002simple}. A necessary assumption of the working model is that the relationship between dose level and dose toxicity is monotonic under a given simple ordering. This allows for multiple models to be specified based on the defined set of simple orderings, with each model corresponding to a specific ordering of doses.

The dose level for the next cohort of patients is allocated by minimising the difference between estimated risk of DLT and the TTR, which can be expressed as the following criterion, \begin{equation} \label{eq:criterion1}
    x_{j+1} = \argmin_{d_{k}} |\widehat{R}(d_{k})- \theta|,
\end{equation} where $\theta$ is the TTR. By repeating this estimation-minimisation process until the stopping conditions are satisfied, an estimate for the true MTD is obtained, which is the dose recommended by the model following the final cohort of patients. 

\subsection{Continual Reassessment Method for Partial Ordering (POCRM)}

Under the Bayesian framework of the POCRM, a prior distribution for the model parameters, $f(a)$, and a prior probability for each ordering $p(m)$ where $\sum\limits_{m=1}^M p(m) = 1$ and $p(m) \geq 0 \; \forall m$ is required. 

Since $Y$ is binary, the likelihood takes the form of a Bernoulli random variable, where each patient in a cohort either experiences or does not experience a DLT. The observed data up to patient $j$ is defined as $\Omega_j = \{x_1, y_1, \ldots, x_j, y_j\}$. This gives the following likelihood under ordering $m$ after the inclusion of $j$ patients in the trial, \begin{equation}
    L_m(a|\Omega_{j})= \prod_{l=1}^{j} \{ \psi_m(x_{l},a) \}^{y_l} \{ 1 - \psi_m(x_l,a) \}^{1 - y_l},
\end{equation}
where $x_l$ is the dose allocated to patient $l$, and $y_l$ is the binary variable denoting whether patient $l$ experiences a DLT and \(\Omega_j\) contains the paired patient data \((x_l,y_l)\). Given the likelihood, the posterior density for the parameter $a$ under the $m$-th model is given by \begin{equation} \label{eq:marginallikelihood}
f_m(a|\Omega_{j}) = \frac{L_m(a|\Omega_j)f(a)}{\int_\mathcal{A} L_m(a|\Omega_j)f(a)da}.
\end{equation}That is, each model will have a unique posterior distribution for $a$, as each model implies a unique ordering of doses. The posterior probabilities for each ordering are also obtained via Bayes' rule as follows, \begin{equation}
    p(m|\Omega_{j})= \frac{p(m) \int_\mathcal{A} L_m(a|\Omega_j)f(a)da}{\sum\limits_{m'=1}^Mp(m')\int_\mathcal{A}L_{m'}(a|\Omega_j)f(a)da}.
\end{equation} The model used for dose allocation is selected by maximising the posterior model probabilities, \begin{equation}
    m^* = \argmax_{m}p(m|\Omega_j),
\end{equation} which results in a single partial ordering being selected for downstream estimation. The posterior density corresponding to this model is then used to estimate a posterior mean for parameter $a$, \begin{equation}
    \widehat{a}_{m^*}= \int_\mathcal{A} a f_{m^*}(a|\Omega_{j})da,
\end{equation} which can be plugged directly into the working model to obtain an estimate for the risk of DLT for the $k$-th dose, \begin{equation}
    \widehat{R}(d_k) = \psi_b(d_k,\widehat{a}_{m^*}),
\end{equation} from which the next dose is allocated using the criterion expressed in Equation (\ref{eq:criterion1}).

\subsection{Bayesian Model Averaging POCRM (BMA-POCRM)}

Suppose that rather than selecting a single model or ordering for each dose allocation in the trial, all orderings are taken into account before making the next decision on dose allocation. Bayesian model averaging (BMA) \citep{raftery1997bayesian} would allow for estimates for probability of toxicity under multiple orderings to be combined for a single probability across several models. This is done by combining the available posterior information on the set of parameters $a$ and the $M$ model probabilities. BMA can be applied to the posterior distribution of the set of parameters $a$. This allows for incorporating uncertainity in $a$ and partial orderings before obtaining an estimate of the risk of DLT of each dose level. 

Using the previously described model-specific posterior distributions of $a$ and the posterior model probabilities $p(m|\Omega_j)$, a weighted combination of posterior distributions is obtained as follows, \begin{equation}
    g(a|\Omega_{j}) = \sum\limits_{m=1}^{M}p(m|\Omega_{j}) f_{m}(a|\Omega_{j}),
\end{equation} where $g(a|\Omega_j)$ is the model averaged posterior of $a$. Hence, rather than having a model-specific posterior for $a$, a version averaged across all models, relative to the posterior probabilities $p(m|\Omega_j)$ is now obtained. 

By applying a change of variables to $g(a|\Omega_j)$ the working model for $R(d_k)$ is expressed as a probability distribution. Although $R(d_k)$ is a useful expression for interpretation, $R(d_k)$ for each $k$ is estimated independently of the others, hence, $R(d_k)$ is not a function of $d_k$ but a function of $a$ since $R(d_k) = \psi_m(d_k, a)$. The change in variable is $R(d_k)=\psi_m(d_k,a)$ such that from the combined posterior distribution the following probability distribution function for $R(d_k)$ can be obtained, \begin{equation}
    |f_m\left(R(d_k|\Omega_j)\right) d R(d_k)| = |f_m(a|\Omega_j) da|,    
\end{equation} since the probability contained under a differential area must be remain constant under a change of variables. Thus, \begin{equation} f_m(R(d_k)|\Omega_j) = \left| \frac{d \psi^{-1}_{m}(d_k, R(d_k))}{d R(d_k)} \right| f_m(a|\Omega_j).\end{equation} Reapplying BMA the following combined posterior distribution for the risk of toxicity is determined, \begin{equation}
    g(R(d_k)|\Omega_j) = \sum_{m=1}^M p(m|\Omega_j) f_m(a|\Omega_j),
\end{equation} which is independent of ordering. The expectation for the risk of toxicity under dose $d_k$ is then, \begin{equation}
    \mathbb{E}[R(d_k)]=\int_{0}^{1}R\left(d_k)\;g_{\psi(d_k,a)}(R(d_k)|\Omega_j\right)\;d\{R(d_k)\}
\end{equation} which under the law of the unconscious statistician \citep{papoulis2002probability} is equivalent to \begin{equation}
    \mathbb{E}[R(d_k)]=\int_\mathcal{A}\psi(d_k,a)g(a|\Omega_j)da,
\end{equation} where \(\mathcal{A}\) is the domain of \(a\). Finally, dose allocation is carried out under this framework by setting $\widehat{R}(d_k)=\mathbb{E}[R(d_k)]$ and applying the criterion described in Section \ref{subsec:generalframework}.

\section{Coherency in the Presence of Partial Ordering}\label{sec:incohstudy}

\subsection{Defining Estimation Coherency}

Coherence in Phase I clinical trials is a useful concept for assessing the theoretical qualities of trial methodology. Throughout this article, coherence, which is defined based on escalation and de-escalation behaviour of a dose-finding model, is referred to as \emph{escalation coherence}. In practice, an escalation coherent design will benefit patient safety as it reduces the likelihood of assigning an overly toxic dose to a patient or cohort but will also ensure that the maximal dose level considered safe is administered. Expanding to the drug combination setting, \citet{park2020coherence} introduces definitions of strong and weak coherency, both of which rely on evaluating the characteristics of escalation and de-escalation to define coherency. A definition of escalation coherency which is equally applicable to both single-agent and dual-agent combination trials is also presented. 

There are two sets of doses that are relevant to escalation coherency. Let \(\mathcal{E}_n\) and \(\mathcal{D}_n\) contain the candidate dose levels for escalation and de-escalation, respectively, for dose allocation \(X_n\). 

\begin{definition}[Escalation Coherency] A design is coherent in dose escalation if $\Pr[X_{n+1} \in \mathcal{E}_n | Y_n = 1] = 0$ for $n=1,\ldots,d-1$ and is coherent in dose de-escalation if $\Pr[X_{n+1} \in \mathcal{D}_n | Y_n = 0] = 0$ for $n=1,\ldots,d-1$. A design is escalation coherent if dose escalation and de-escalation are coherent.
    
\end{definition} 

This definition emphasises the sole concept of dose selection. Whereas, during the administration of a real-world trial, there is frequent interface between domain expert and dose-escalation model. Domain experts use both the given toxicity estimates and the recommended next dose to make a final escalation decision. This is exceedingly relevant to the combination setting where there could be several combinations to choose from at any given point in the trial. To address the need for both toxicity estimates and dose escalation recommendations to be coherent, the following specification of \emph{estimation coherency} is introduced.

\begin{definition}[Estimation Coherency] \label{def:estimation-incoh}
Suppose there is a given partial ordering, from which $M$ simple orderings are derived. Let $I_m(d_i)$ be the index of dose $i$ within ordering $m$. If dose $i$ is more toxic than dose $j$ under ordering $m$ then $I_m(d_j) < I_m(d_i)$ and vice versa. Based on the $M$ orderings, each dose $i$ will have an \textit{a priori} known set of less toxic doses, \[\nu_i = \{d_j : I_m(d_j) < I_m(d_i) \; \forall m\},\] and similarly a known set of more toxic doses \[\xi_i = \{d_j:I_m(d_j) > I_m(d_i) \; \forall m\}.\] A design is estimation coherent if it satisfies both (i)  following no DLT at \(d_i\) the estimated risk of toxicity for dose $j$, $\widehat{R}(d_j)$, decreases for all \(d_j \in \nu_i \cup \xi_i\), and (ii) following DLT at \(d_i\), the estimated risk of toxicity for $d_j$, \(\widehat{R}(d_j)\), increases for all \(d_j \in \nu_i \cup \xi_i\). 
\end{definition}

This definition of estimation coherency acts as a 'sanity check' for the dose-finding model. By ensuring the model coincides with both the prior knowledge implied by the partial ordering and most recently gained information, it is able to detect changes which are potentially illogical and may endanger the integrity of the trial. This type of coherency is not relevant to single dose-finding methods as, under monotonicity, this condition is always satisfied.

\subsection{Illustrative Example} \label{subsec:incohsetting}

The POCRM is particularly vulnerable to estimation incoherencies. Here, a specific example of estimation incoherency under the POCRM is explored.
 
Consider the $3 \times 2$ drug combination design shown in Table \ref{tab:casestudy}. Under this combination setting, the chosen cohort size is 1 with a TTR of 0.4. Applying the partial ordering specification recommended by \citet{wages2013specifications}, the following 6 simple orderings are obtained, of which 5 are unique,

\begin{align*}
    m=1: \; d_1 \rightarrow d_2 \rightarrow d_3 \rightarrow d_4 \rightarrow d_5 \rightarrow d_6, \\ 
    m=2: \; d_1 \rightarrow d_3 \rightarrow d_5 \rightarrow d_2 \rightarrow d_4 \rightarrow d_6, \\ 
    m=3: \; d_1 \rightarrow d_3 \rightarrow d_2 \rightarrow d_5 \rightarrow d_4 \rightarrow d_6, \\
    m=4: \; d_1 \rightarrow d_2 \rightarrow d_3 \rightarrow d_4 \rightarrow d_5 \rightarrow d_6, \\
    m=5: \; d_1 \rightarrow d_2 \rightarrow d_3 \rightarrow d_5 \rightarrow d_4 \rightarrow d_6, \\
    m=6: \; d_1 \rightarrow d_3 \rightarrow d_2 \rightarrow d_4 \rightarrow d_5 \rightarrow d_6.
\end{align*} This set of orderings is the complete set of possible orderings, assuming that dose-toxicities increase monotonically only where the dose level increases in only one drug of the combination. At the start of the trial, the \emph{a priori} probability for each ordering is equal. 

From these simple orderings one can obtain the sets of interest \(\nu_i\) and \(\xi_i\) for each dose as shown for each dose level in Table \ref{ei_sets}. The sets used to check for estimation incoherencies are composed as follows. For \(d_2\), $\nu_2 = \{d_1\}$ and $\xi_2 = \{d_4, d_6\}$. This is obtained by considering that \(d_1\) is less toxic than \(d_2\) under every simple ordering, and both \(d_4\) and \(d_6\) are more toxic than \(d_2\) under every simple ordering. These sets exist for each dose level and can be used following each Bayesian probability update to detect any estimation incoherencies.  

\begin{table}
\centering
\begin{tabular}{@{}ccc@{}}
\toprule
Dose Level, $i$ & $\nu_i$                & $\xi_i$       \\ \midrule
1               & $\emptyset$                   & $\{d_2, d_3, d_4, d_5, d_6\}$ \\
2               & $\{d_1\}$                     & $\{d_4, d_6\}$                \\
3               & $\{d_1\}$                     & $\{d_4, d_5, d_6\}$           \\
4               & $\{d_1, d_2, d_3\}$           & $\{d_6\}$                     \\
5               & $\{d_1, d_3\}$                & $\{d_6\}$                     \\
6               & $\{d_1, d_2, d_3, d_4, d_5\}$ & $\emptyset$                   \\ \bottomrule
\end{tabular} 
\caption{Sets for detecting estimation incoherencies where $\nu_i$ is the set of doses always less toxic than dose $i$ and $\xi_i$ is the set of doses always more toxic than dose $i$. \label{ei_sets}}
\end{table} 

\subsection{Applying POCRM with Model Selection}

The parametrisation of POCRM used here matches that which is used in the full simulation study as described in Section SM.2 of the Supplementary Materials with a cohort size of 1. 1,000 simulations were initiated with the first estimation incoherency identified in the second simulated trial at the induction of cohort 12. At this point in the trial, the posterior ordering probabilities are shown in Table \ref{tab:completeincohstudy}. The dose allocations and DLTs observed up to and including cohort 11 are as follows, \begin{align*} \mathbf{n}^{[11]} &= (1, 0, 1, 6, 2, 1), \\ 
\mathbf{y}^{[11]} &= (0, 0, 0, 3, 1, 1), \end{align*} where the $i$-th entry in \(\mathbf{n} \in \mathbb{Z}^6\) is the number of patients assigned to the $i$-th dose level, $d_i$, and \(\mathbf{y} \in \mathbb{Z}^6\) is the number of patients that experienced a DLT after being assigned dose level $d_i$. As seen in Table \ref{tab:completeincohstudy}, following the induction of cohort 11, $d_2$ is recommended as the next dose by POCRM. Hence, cohort 12 is inducted and administered $d_2$, which yields the following allocation vectors, \begin{align*} \mathbf{n}^{[12]} &= (1, 1, 1, 6, 2, 1), \\ 
\mathbf{y}^{[12]} &= (0, 0, 0, 3, 1, 1), \end{align*} where no new DLT is observed for $d_2$. 

Recalling the set of doses with known toxicity relative to $d_2$ in Section \ref{subsec:incohsetting}, since no DLT is observed at $d_2$ the dose-toxicity estimates for all dose levels in $\nu_2 \cup \xi_2 = \{d_1, d_4, d_6\}$ are expected to decrease. However, in Table \ref{tab:completeincohstudy} an increase in the toxicity estimate for $d_4$ from 0.49 to 0.56 is observed. This is a change of +0.07 despite there being no information gained from the previous cohort that indicates a greater toxicity of $d_4$. Throughout the coming analyses, the occurrence of estimation incoherencies is considered as a key operating characteristic of the methods being studied. 

\captionsetup{width=\textwidth}

\begin{table}[]
\centering
\begin{tabular}{@{}cccccccc@{}}
\toprule
\multirow{2}{*}{Method} & \multirow{2}{*}{Cohort} & \multicolumn{6}{c}{\(p(\cdot|x):\) (Posterior Model Probability)}     \\
                           &    & \(m=1\) & \(m=2\) & \(m=3\) & \(m=4\)         & \(m=5\)      & \(m=6\)      \\ \midrule
\multirow{2}{*}{POCRM}     & 11 & 0.1568  & 0.1497  & 0.1878  & 0.1568          & 0.1582       & {\ul 0.1906} \\
                           & 12 & 0.1743 & 0.1091 & 0.1840 & 0.1743         & {\ul 0.1840} & 0.1743      \\ \midrule
\multirow{2}{*}{BMA-POCRM} & 11 & 0.1568  & 0.1497  & 0.1878  & 0.1568          & 0.1582       & {0.1906}       \\
                           & 12 & 0.1743  & 0.1091  & 0.1840  & 0.1743          & { 0.1840}       & 0.1743       \\ \midrule \midrule
\multirow{2}{*}{Method} & \multirow{2}{*}{Cohort} & \multicolumn{6}{c}{\(R(\cdot):\) (Estimated Probability of Toxicity)} \\
                           &    & \(d_1\) & \(d_2\) & \(d_3\) & \(d_4\)         & \(d_5\)      & \(d_6\)      \\ \midrule
\multirow{2}{*}{POCRM}  & 11                      & 0.0672  & {\ul 0.3261}  & 0.1756  & \textcolor[HTML]{F44336}{\textbf{0.4859}} & 0.6282 & 0.7412 \\
                           & 12 & 0.0331  & 0.1114  & 0.2432  & \textcolor[HTML]{F44336}{\textbf{0.5562}} & {\ul 0.4023} & 0.6853       \\ \midrule
\multirow{2}{*}{BMA-POCRM} & 11 & 0.0802  & 0.2671  & 0.2371  & 0.5247          & {\ul 0.5019} & 0.7111       \\
                           & 12 & 0.0654  & 0.2281  & 0.2210  & 0.4975          & {\ul 0.4860} & 0.6933       \\ \bottomrule
\end{tabular}
\caption{Model probabilities and dose-toxicity estimates for POCRM and BMA-POCRM dose-escalation frameworks in a setting with a TTR of 0.4. The recommended dose levels and selected orderings following each cohort are shown with an underline. Incoherencies are shown in red. \textbf{Note:} Despite, the recommended dose-level given by BMA-POCRM being $d_5$ following cohort 11, $d_2$ is treated as the recommended dose to maintain consistency for comparison of POCRM and BMA-POCRM.}
\label{tab:completeincohstudy}
\end{table}

\captionsetup{width=0.9\textwidth}

This example illustrates the often illogical changes in toxicity estimates observed under the POCRM. Since toxicity estimates guide clinicians and affect the next allocated dose, it is crucial that changes in these toxicity estimates are robust to scrutiny. 

\subsection{Applying BMA-POCRM} \label{sec:application}

Again, under this setting, the dose-toxicity estimates for all dose levels in $\nu_2 \cup \xi_2 = \{d_1, d_4, d_6\}$ are expected to decrease or remain the same as there was no DLT observed for $d_2$. Returning to the results presented in Table \ref{tab:completeincohstudy} for BMA-POCRM, for all doses in $\nu_2 \cup \xi_2$ the toxicity estimate decreases. Conversely to POCRM, BMA-POCRM remains coherent with respect to estimation coherency in this case. 

\section{Case Study} \label{sec:realtrial}

To consider the performance of BMA-POCRM in comparison to POCRM in a real-world setting, we apply these methods to a Phase I study dosing patients combinations of neratinib and temsirolimus \citep{gandhi2014phase}. This trial involved fifty-two patients treated on 12 doses in a 4-by-4 grid of possible neratinib-temsirolimus combinations with a TTR of \(\theta=1/3\). The remaining 4 doses in the grid were never assigned to patients in the original trial. A DLT is defined as an inability to maintain the prescribed dose for the first 28 days of treatment due to treatment-related toxicity. The two initial cohorts consisted of two patients each were enrolled simultaneously with: (1) 160 mg of neratinib/15 mg of temsirolimus; and (2) 120 mg of neratinib/25mg of temsirolimus. 

We apply BMA-POCRM and POCRM the parameters selected in the calibration study in Section SM.2 and follow the approach of \cite{wages2013specifications} to derive candidate orderings based on the 12 available doses in the trial.  The full trial data used for these simulations can be found in Table S3 in the Supplementary Materials.

The aim here is to identify incoherencies and large changes in toxicity estimates. By showing that these occur in real-world trials we aim to further motivate the use of BMA-POCRM in practice. 

\subsection{Data Generation}


To allow for a fair evaluation of model-guided dose escalation, we use the original trial data to conduct a case study according to the scheme outlined by \cite{barnett2021comparison}. We define a fixed set of fifty-two patient dose responses for each dose. Here, we denote the number of patients assigned to dose \(j\) by $n_j$ and the number of observed DLTs under dose \(j\) by \(y_j\).

To define a fixed set of fifty-two patient dose responses for each dose we take the first \(n_j\) responses to be a random permutation of responses from the original study. The remaining \(52 - n_{j}\) responses are generated from a Beta$(1 + y_{j}, 1 + n_{j} - y_{j})$ distribution by first sampling a probability of DLT and sampling a binary response from a Bernoulli distribution with the given probability of DLT. Where no patients are assigned to a dose combination in the real study, probabilities are generated from a Beta$(3,3)$ distribution instead. As each method allocates a dose to each cohort, the ordered set of responses corresponding to each dose is used to determine treatment response sequentially. This ensures that the $n$th patient allocated to each dose, regardless of method of dose escalation used, will have the same response to treatment.

\subsection{Results}

\begin{figure}[h!]
    \centering
    \includegraphics[
   width=0.5\textwidth] {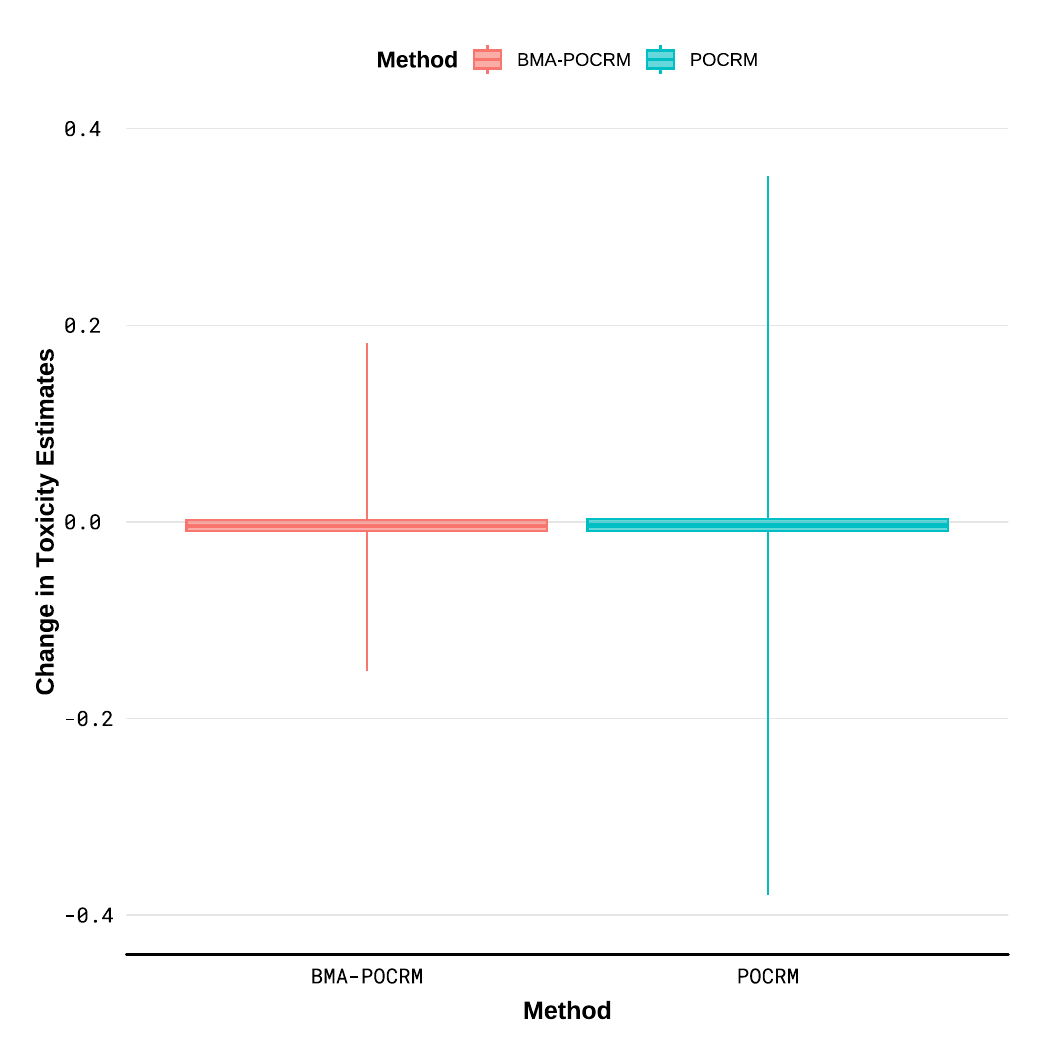}
    \caption{Distribution of the magnitude of changes in toxicity estimates in the motivating trial simulation. Whiskers indicate range of maximum and minimum values observed.}
    \label{fig:casestudy_mag}
\end{figure}

Figure \ref{fig:casestudy_mag} shows that the range of magnitude of changes in toxicity estimate for POCRM is larger than for BMA-POCRM. The range is shown to be [-0.37, 0.35] and [-0.17, 0.18] for POCRM and BMA-POCRM, respectively. 

\begin{figure}
    \centering
    \includegraphics[width=0.75\textwidth]{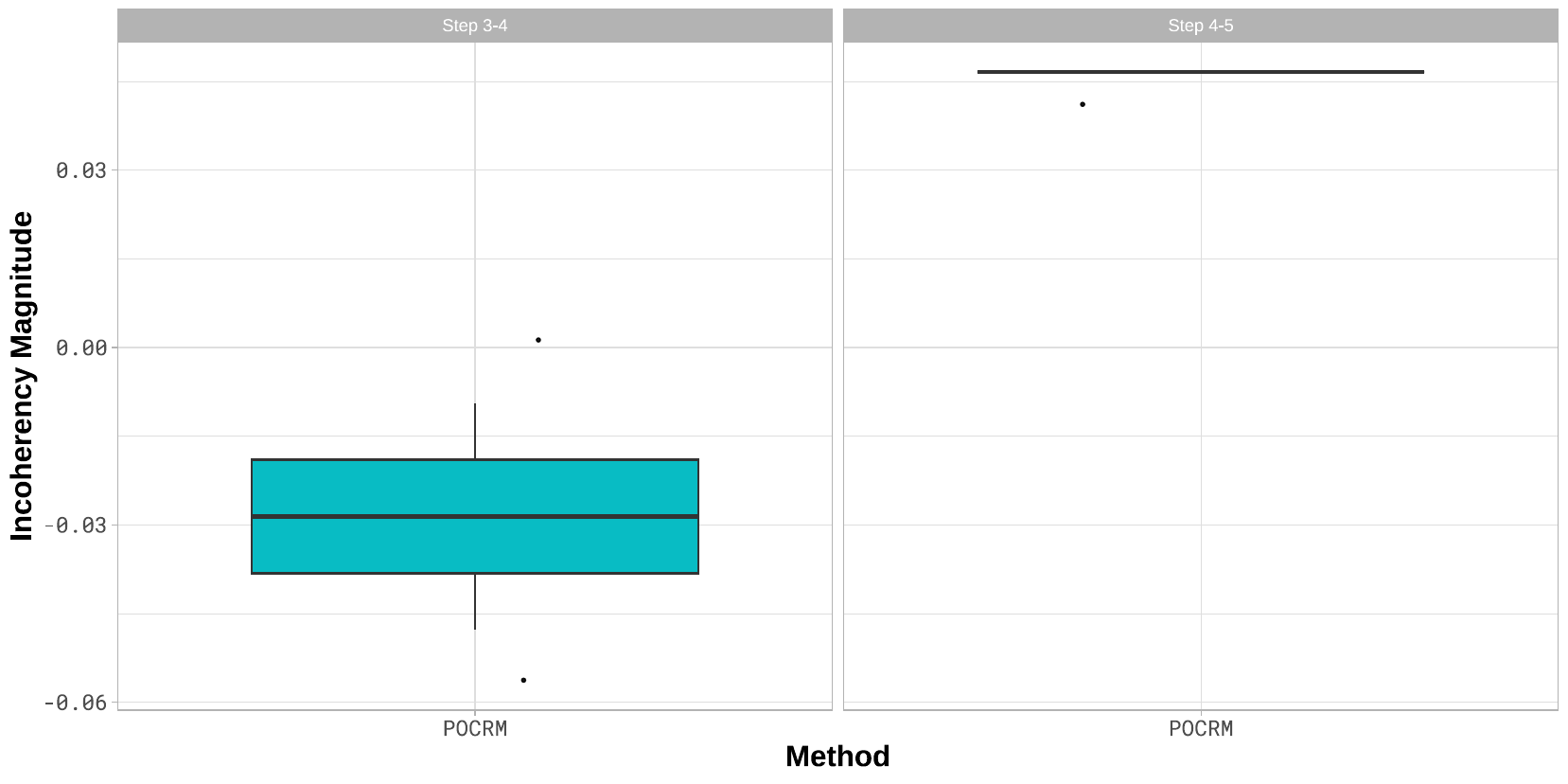}
    \caption{Incoherencies that occurred in motivating trial.}
    \label{fig:incoh_motivating}
\end{figure}

Table \ref{tab:motivating_jumps} shows an example of a cohort where the large change in toxicity estimate is particularly great. After assigning $d_8$ to cohort 3, we observe a DLT, which leads to several incoherencies and large changes in toxicity estimates for the estimates that are to be used to assign doses to cohort 4. For POCRM, we observe two changes that are $>25$, for $d_4$ and the administered dose $d_8$. We also observe two estimate incoherencies for doses $d_5$ and $d_6$, which we know are less toxic than $d_8$ according to the partial orderings corresponding to this trial setting. For this cohort, the toxicity estimates given by BMA-POCRM do not yield any incoherencies or large changes as is true for the remainder of the motivating trial, as shown in Figure \ref{fig:incoh_motivating}.

\begin{table}[h!]
\centering
\begin{adjustbox}{width=\textwidth, totalheight=0.9\textheight, keepaspectratio}
\begin{tabular}{cccccccccccccccc}
\toprule
                            &                          & \multicolumn{12}{c}{$R(\cdot):$ (Estimated Probability of Toxicity)} & \multirow{2}{*}{$d_\text{sel}$} & \multirow{2}{*}{$y_\text{sel}$}                                  \\
\multirow{-2}{*}{Method}    & \multirow{-2}{*}{Cohort} & $d_1$ & $d_2$ & $d_3$ & $d_4$ & $d_5$ & $d_6$ & $d_7$ & $d_8$ & $d_9$ & $d_{10}$ & $d_{11}$ & $d_{12}$                      \\ \midrule
                            & 3                        & 0.00 &  0.00 & 0.01 & {\color[HTML]{F44336} \textbf{0.04}} & {\color[HTML]{2F3FEB} \textbf{0.08}} & {\color[HTML]{2F3FEB} \textbf{0.15}} & 0.23 &   {\color[HTML]{F44336} \textbf{0.33}} & 0.43 & 0.53 & 0.61 & 0.69 & $d_8$ & 1 \\
\multirow{-2}{*}{POCRM}     & 4                        &  0.00 & 0.01 & 0.07 & {\color[HTML]{F44336} \textbf{0.32}} & {\color[HTML]{2F3FEB} \textbf{0.03}} & {\color[HTML]{2F3FEB} \textbf{0.14}} & 0.42 & {\color[HTML]{F44336} \textbf{0.68}} & 0.22 & 0.51 & 0.75 & 0.60 & -- & --  \\ \midrule
                            & 3                        & 0.01 & 0.04 & 0.13 & 0.25 & 0.06 & 0.13 & 0.29 & 0.48 & 0.18 & 0.36 & 0.52 & 0.41 & $d_{10}$ & 0
                            \\
\multirow{-2}{*}{BMA-POCRM} & 4                        & 0.01 & 0.02 & 0.09 & 0.20 & 0.03 & 0.08 & 0.22 & 0.41 & 0.11 & 0.27 & 0.44 & 0.32 & -- & --                        \\ \bottomrule
\end{tabular}
\end{adjustbox}
\caption{Example of a sudden change in toxicity estimates under POCRM in the motivating trial. Any sudden changes (i.e. those greater than 0.25) are shown in {\color[HTML]{F44336} \textbf{red}} and any incoherencies are shown in {\color[HTML]{2F3FEB} \textbf{blue}}.}
\label{tab:motivating_jumps}
\end{table}

Figure \ref{fig:incoh_motivating} shows that incoherencies only occurred for POCRM in the motivating trial. The incoherencies shown for step 3-4 correspond to the results in Table \ref{tab:motivating_jumps}. The full dose allocations can be found in the Appendix in Tables S4 and S5.

\section{Simulation Study} \label{sec:simstudy}

\subsection{Specification} \label{sec:workingmodel}

A simulation study is necessary to evaluate and compare the operating characteristics for POCRM and BMA-POCRM in a large-scale setting. For this study, a 4-by-4 structure is selected in a 2 drug combination study. For each scenario, $10,000$ trial simulations are run with a TTR of $\theta = 0.3$ and a cohort size of $1$ for $60$ cohorts. In Section \ref{sec:sensitivity}, cohort sizes of 3 and 6 are considered with 20 and 10 cohorts, respectively. We select model parameters based on the calibration study presented in Section SM.2.

There are several candidates for a working model $\psi_m(d_k, a)$. The parametrisation of the power model given in \citet{wages2011continual} is used here and takes the following form, $$\psi_m(d_k, a) = \alpha_{mk}^a, \;\;\; \; k = 1, \ldots, K$$ where $a \in [0, \infty)$ and $0 < \alpha_{m1} < \ldots, < \alpha_{mK}$ is the probability skeleton, which represents the prior estimates of dose toxicity at each dose level under ordering $m$. The prior distribution of $a$ is Normal with mean 0 and variance 1.34 as suggested by \citet{wages2011continual}. The selected variance of the normal prior has no impact on model performance as shown in SM.3. This working model is applied throughout the paper for implementations of both POCRM and BMA-POCRM.

Five metrics are used to assess models in simulation trials according to recommendation accuracy, assignment accuracy and patient safety. In this case, an overly toxic dose is chosen to be any dose that has a true probability of toxicity greater than $110\%$  of the TTR. \begin{enumerate}[(i)]
    \item The proportion of correct selections (PCS) is the proportion of trials that recommended doses with a true probability of toxicity equal to the TTR.
    \item The proportion of acceptable selections (PAS) is the proportion of trials that recommended a dose with a true probability of toxicity within $[\theta - 0.1, \theta]$ where $\theta$ is the TTR.
    \item Proportion of trials that give overly toxic selections (POTS).
    \item The number of patients treated at overly toxic doses (NPTOT).
    \item Finally, the estimation coherency of model estimates at the induction of each trial as defined in Definition \ref{def:estimation-incoh}.
\end{enumerate} 
 
The full specification for all 24 scenarios can be found in Table S1 of the Supplementary Materials. Scenarios with both symmetric and asymmetric true MTD locations are included. The number of accurate orderings present in the set of candidate orderings is also varied across scenarios. Those found in the left-most column of Table S1 have no correct orderings, implying that none of the orderings exactly match the specified probabilities of toxicity. Scenarios found in the middle column are satisfied by a single ordering in the list of candidate orderings. Finally, those found in the right-most column are equally satisfied by 2 candidate orderings.

\captionsetup{width=0.9\textwidth}

\subsection{Results} \label{sec:simresults}

\begin{figure}[h!]
    \centering
    \includegraphics[width = 0.75\textwidth]{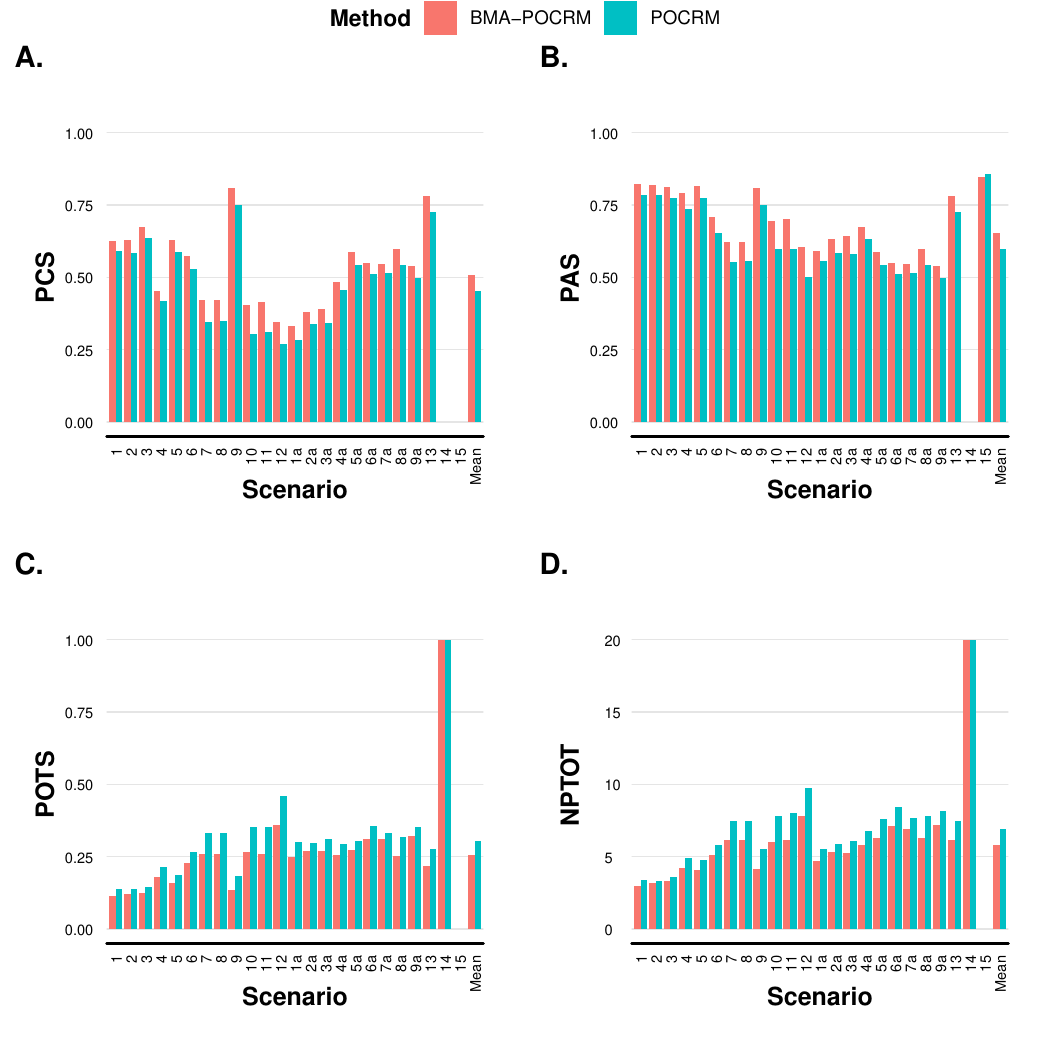}
    \caption{Panel A. Arithmetic mean PCS across 10,000 repeated simulations with a cohort size 1 for each scenario. The mean performance across scenarios is also shown. Panel B. Arithmetic mean PAS across 10,000 repeated simulations with a cohort size 1 for each scenario. The mean performance across scenarios is also shown. Panel C. Arithmetic mean POTS across 10,000 repeated simulations with a cohort size 1 for each scenario. The mean performance across scenarios is also shown. Panel D. Arithmetic mean PTOTD across 10,000 repeated simulations with a cohort size 1 for each scenario. The mean performance across scenarios is also shown.}
    \label{fig:sim_plot_coh1}
\end{figure}

Figure \ref{fig:sim_plot_coh1} illustrates model performance under all $24$ scenarios, and provides an arithmetic mean performance across scenarios. BMA-POCRM exhibits better PCS and PAS across all scenarios but scenario 15, indicating that it more consistently selects desirable dose levels following the Phase I trial simulations. 

BMA-POCRM leads to an average increase of 5.2\% in PCS when compared to that of POCRM. Across all scenarios with non-zero PCS, there is an improvement in performance when applying BMA-POCRM. Considering scenarios 1-13, no scenario shows a lower difference than 2.77\% in favour of BMA-POCRM. Scenario 12 shows the greatest discrepancy between the methods with BMA-POCRM exceeding POCRM by 10.7\%. Likewise, investigating Figure \ref{fig:sim_plot_coh1}.B, BMA-POCRM leads to an improvement by at least 3.34\% and an average of 5.5\%. For this measure, scenario 12 leads to the greatest difference in performance with 10.5\%. Differences in the number of correct orderings between scenarios does not clearly favour either method and there is no consistent effect on performance associated with changes in the number of correct orderings. 

The design matrices for scenarios 14 and 15 are such that there are only overly toxic doses and doses with a lower toxicity than the TTR, respectively. Under scenario 15, POCRM has a mean PAS of $85.61\%$ whilst BMA-POCRM has one of $84.67\%$. Nearly half of the doses in scenario 15 are considered acceptable by definition. Noting that these doses are in the lower triangular of the combination matrix, these are the most toxic doses under both drugs. This suggests that BMA-POCRM is the more conservative approach to dose escalation. 

The safety of BMA-POCRM is further supported by Figures \ref{fig:sim_plot_coh1}.C and \ref{fig:sim_plot_coh1}.D, where it recommends overly toxic doses in a smaller proportion of trials and allocates overly toxic doses to fewer patients, respectively. This trend is maintained throughout all scenarios as the mean POTS is 4.89\% lower for BMA-POCRM than POCRM. Scenario 2 exhibits the smallest difference between methods in terms of POTS with a 1.84\% decrease in the number of overly toxic selections with BMA-POCRM. In scenario 12, BMA-POCRM leads to a 10\% reduction in the number of overly toxic doses. Figure \ref{fig:sim_plot_coh1}.D shows that on average, at least one cohort less will be assigned an overly toxic dose. 

\begin{figure}[h!]
    \centering
    \includegraphics[width=0.75\textwidth]{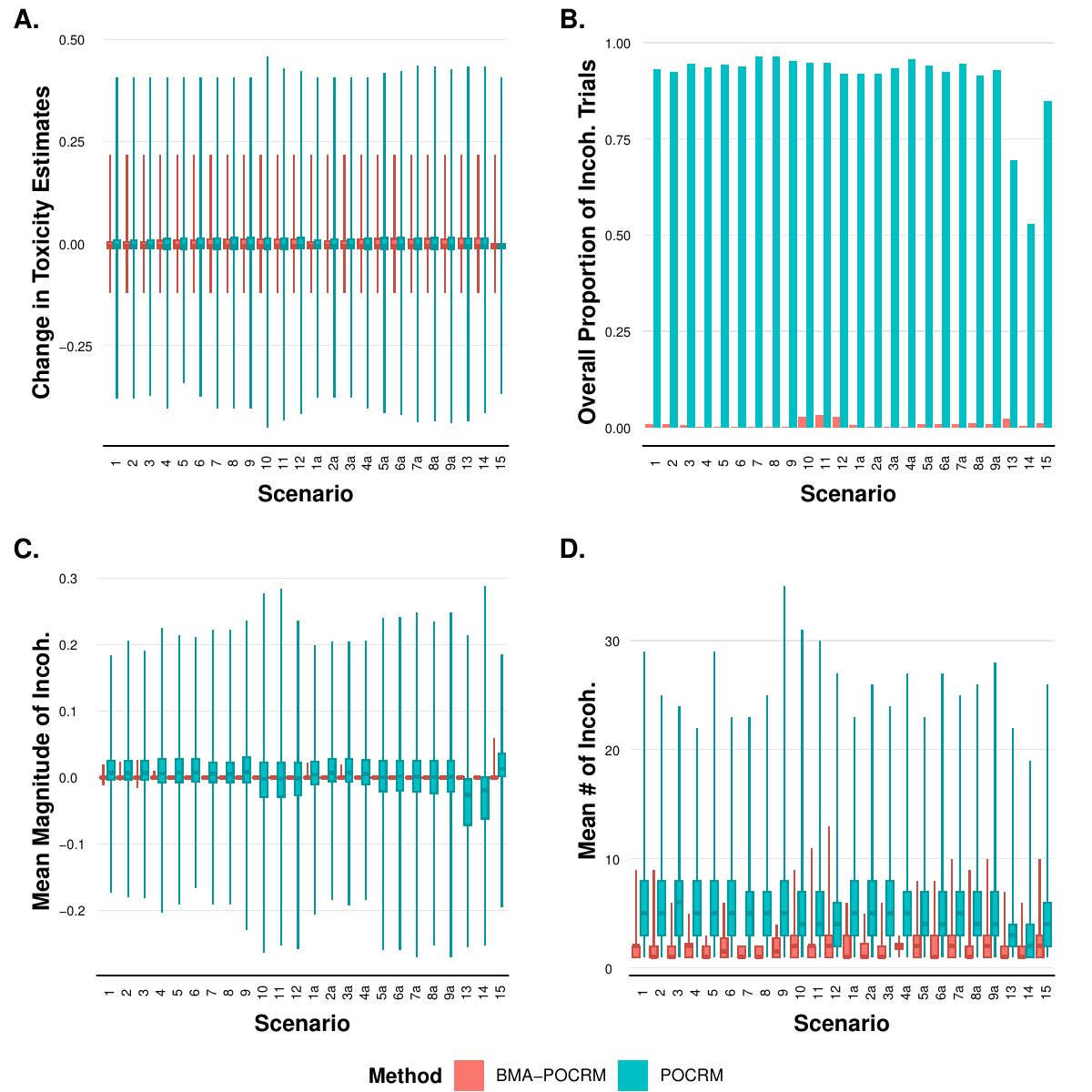}
    \caption{Panel A. Distribution of changes in dose toxicity estimates following the induction of each cohort for simulated trials with cohort size 1. Panel B. Proportion of trials where at least one incoherent estimate is observed. Panel C. Magnitude of incoherent changes in toxicity estimates. Panel D. Mean number of cohorts with at least one observed incoherency.}
    \label{fig:incoh_plot}
\end{figure}

\begin{figure}[h!]
    \centering
    \includegraphics[width=0.75\textwidth]{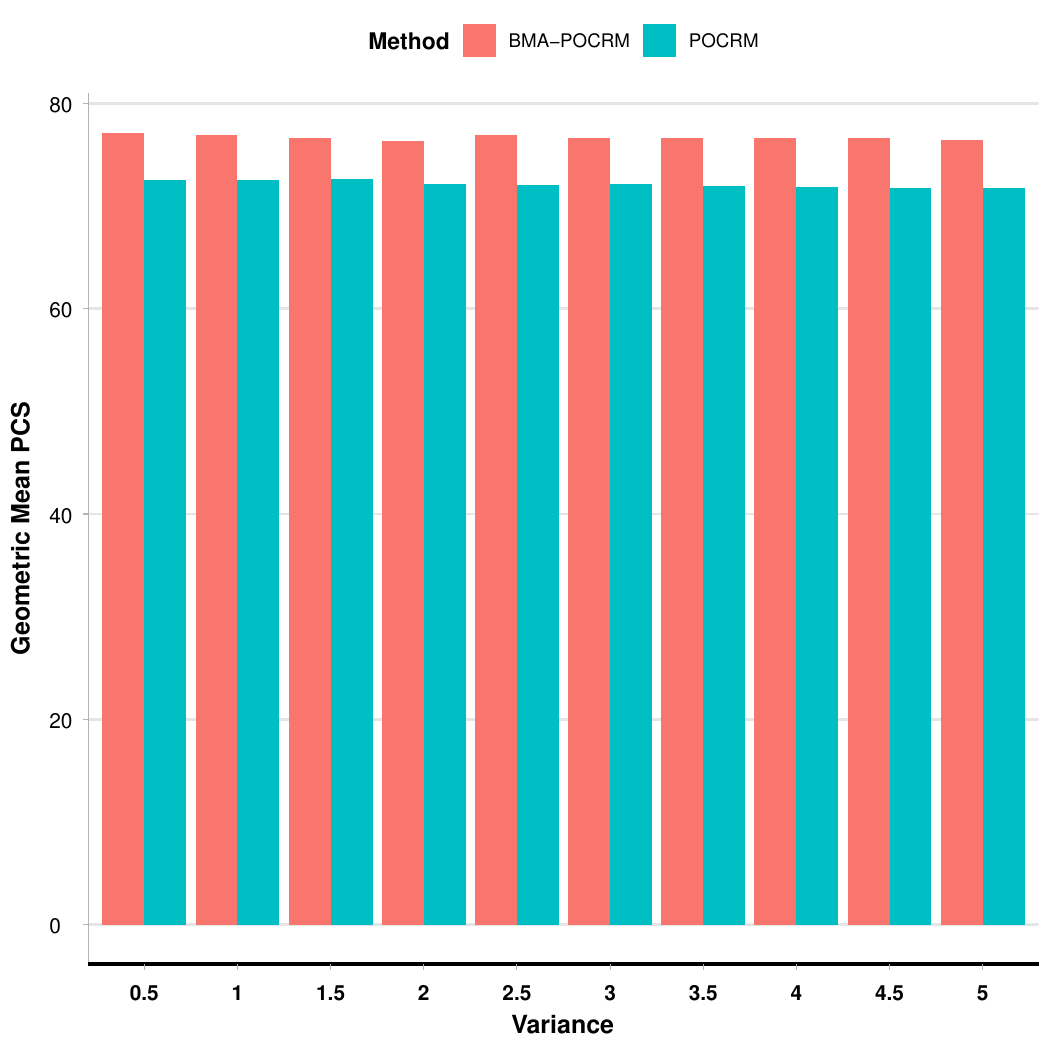}
    \caption{Geometric mean of PCS across all scenarios. The optimal skeleton, $\mathcal{S}$, is used here.}
    \label{fig:var_calib}
\end{figure}

Figure \ref{fig:incoh_plot}.A shows the magnitude and direction of changes in toxicity estimates across all trial steps in the simulated trials. From \ref{fig:incoh_plot}.A, BMA-POCRM has less dispersed both in terms of the box and whiskers for each scenario. Conversely, for changes in model probabilities shown in Figure \ref{fig:var_calib}, since the scheme for obtaining posterior model probabilities is shared between BMA-POCRM and POCRM, the distributions are similar.

The frequency of incoherencies within simulated trials is shown in Figure \ref{fig:incoh_plot}.B. The previous definition of estimation incoherency is here applied to evaluate the estimation operating characteristics of POCRM and BMA-POCRM. In all scenarios but 13, 14 and 15, POCRM leads to at least $90\%$ of trials with at least one estimation incoherent change of toxicity estimates during the trial. Meanwhile, BMA-POCRM exhibits estimation incoherencies in a worst-case $3.13\%$ of trials under scenario 11 and no incoherencies in most scenarios. Further, examining Figure \ref{fig:incoh_plot}.C under scenario 11, the magnitude of estimation incoherent toxicity estimates is significantly smaller for BMA-POCRM with a maximum incoherent change in toxicity estimate of $0.060$ compared to POCRM for which this is $0.288$. Figure \ref{fig:incoh_plot}.D shows that in trials where BMA-POCRM does exhibit estimation incoherencies, this occurs for a smaller number of cohorts. The distribution of the number of cohorts where BMA-POCRM exhibits an incoherency is consistently biased towards zero when compared to those of POCRM.

\section{Sensitivity Analysis}\label{sec:sensitivity}

The operating characteristics of BMA-POCRM illustrated in the previous section show that for a cohort size of 1, this method outperforms the existing POCRM. In this section, the analysis conducted for cohorts of 1, are expanded to cohort sizes of 3 and 6. The simulation settings remain the same as those described in Section \ref{sec:simstudy}. Moreover, the calibrated designs selected in Figure S2 are also used. With a cohort size of 3 and 6, the number of cohorts falls to 20 and 10, respectively, to maintain a constant number of patients at trial completion.

\begin{figure}[h!]
    \centering
    \includegraphics[width=0.75\textwidth]{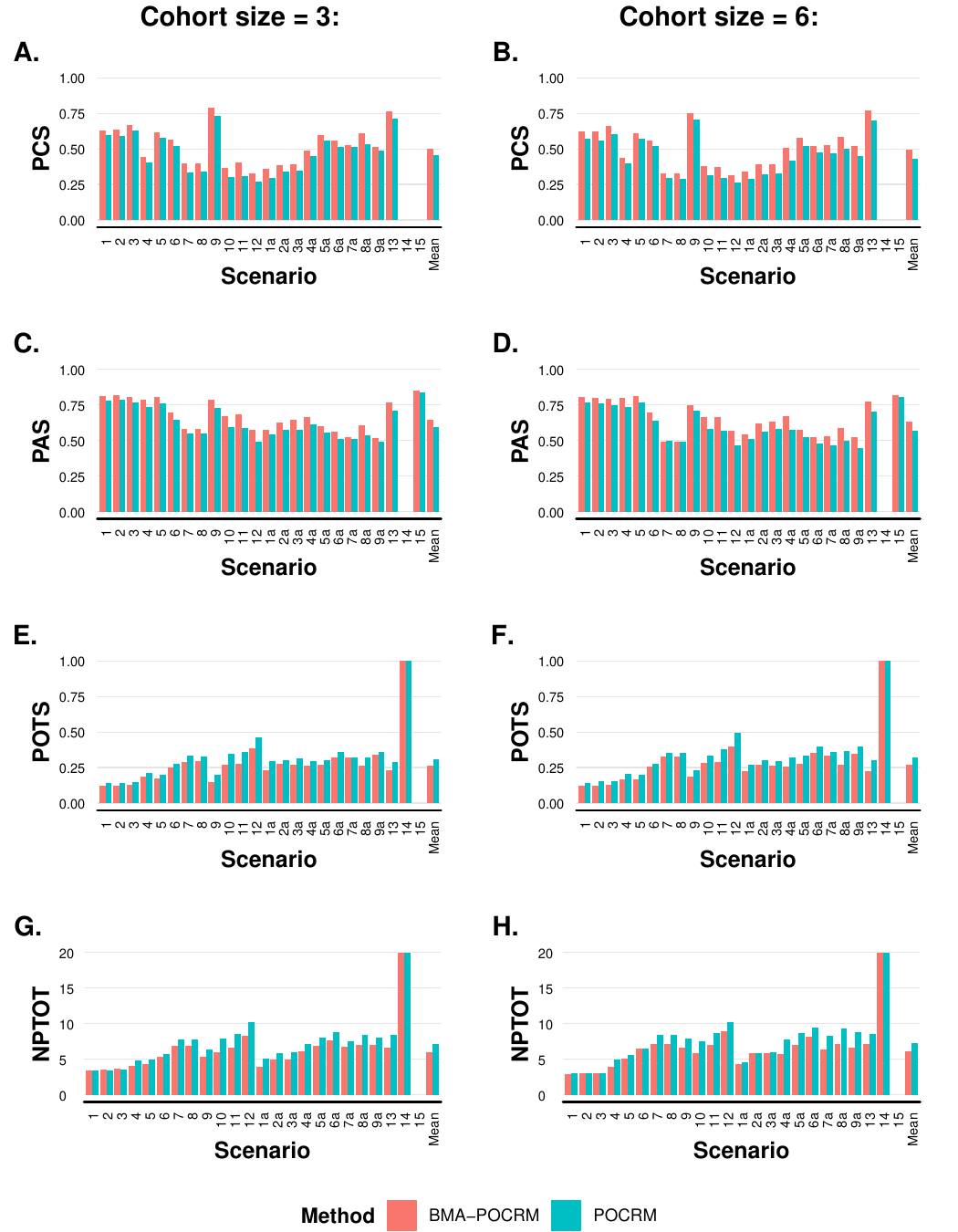}
    \caption{Panel A. Geometric mean of PCS for simulated trials with cohort size 3 and 20 cohorts. 
    Panel B. Geometric mean of PCS for simulated trials with cohort size 6 and 10 cohorts. 
    Panel C. Geometric mean of PAS for simulated trials with cohort size 3 and 20 cohorts.
    Panel D. Geometric mean of PAS for simulated trials with cohort size 6 and 10 cohorts.    
    Panel E. Geometric mean of PTOS for simulated trials with cohort size 3 and 20 cohorts. 
    Panel F. Geometric mean of PTOS for simulated trials with cohort size 6 and 10 cohorts.
    Panel G. Geometric mean of NPTOT for simulated trials with cohort size 3 and 20 cohorts. Panel H. Geometric mean of NPTOT for simulated trials with cohort size 6 and 10 cohorts.}
    \label{fig:sensitivity_analysis}
\end{figure}

Figure \ref{fig:sensitivity_analysis} shows that there is no significant change in performance when increasing cohort size to 3. Specifically, in Figure \ref{fig:sensitivity_analysis}.A-B, BMA-POCRM has mean PCS of 50.42\% compared to that of POCRM of 45.41\%, a difference of 5.01\%. For PAS, mean performance was 64.34\% and 59.12\% for BMA-POCRM and POCRM respectively, a difference of 5.25\%. These discrepancies are consistent with those observed for a cohort size of 1. Furthermore, when further increasing cohort size to 6, there is a gap of 6.44\% and 6.59\% in mean PCS and mean PAS, respectively. For both cohort sizes 3 and 6 under scenario 15, where POCRM previously had superior PAS, BMA-POCRM is now dominant. 

The gains in trial safety also remain consistent with previous simulations as shown in Figures \ref{fig:sensitivity_analysis}.E-G. BMA-POCRM recommends fewer overly toxic doses and a smaller number of patients are exposed to overly toxic doses relative to POCRM. For a cohort size of 3, the greatest improvement in safety is seen for scenario 11 where 8.66\% of trials result in an overly toxic dose being selected. Across scenarios, the average difference in POTS between methods was 4.3\%. For scenario 22, 5.6 fewer patients were exposed to overly toxic doses as a result of BMA-POCRM. Under a cohort size of 6, a mean decrease in POTS of 5.4\% is observed with BMA. The greatest difference in POTS is observed for scenario 8A where BMA-POCRM selects an overly toxic dose in 10.1\% fewer trials. The average number of patients exposed to overly toxic doses also decreases from 21.9 to 18.4, with the sharpest difference being for scenario 9A with 6.5 fewer patients exposed to overly toxic doses. The rate of overly toxic dose assignment was most similar under scenario 1 where there was only a difference of 0.1 patients. 

\begin{figure}[h!]
    \centering
    \includegraphics[width=0.75\textwidth]{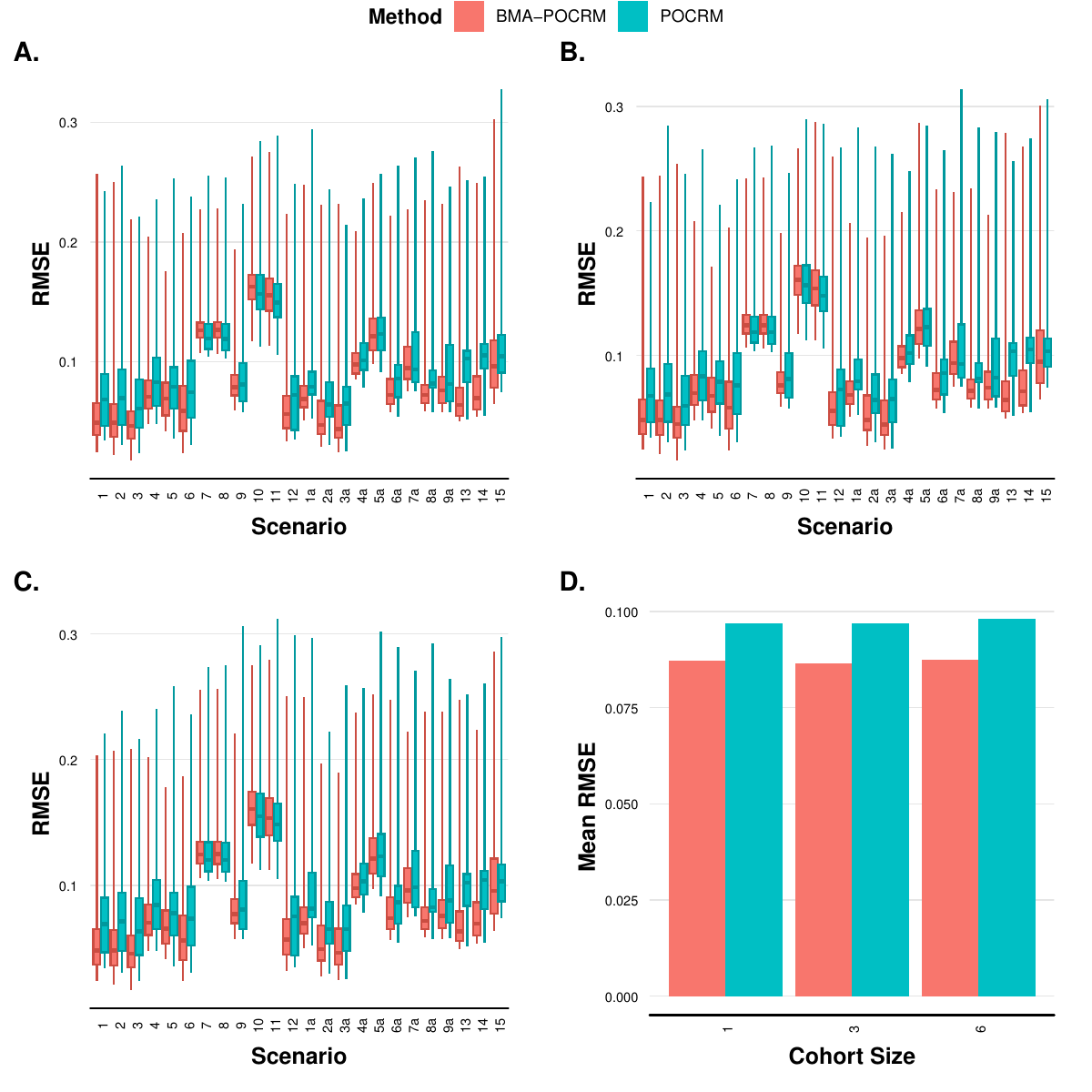}
    \caption{Panel A. Distribution of RMSE at the end of a simulated trial for cohort size of 1. Panel B. Distribution of RMSE at the end of a simulated trial for cohort size of 3. Panel C. Distribution of RMSE at the end of a simulated trial for cohort size of 6. Panel D. Mean RMSE across scenarios for all cohort sizes.}
    \label{fig:rmse}
\end{figure}

Figure \ref{fig:rmse}.A shows the root mean square error (RMSE) of toxicity estimates at the conclusion of trials for a cohort size of 3. In 19 out of 24 scenarios, BMA-POCRM has toxicity estimates closer to the true values than POCRM. This is most exagerated for scenario 13 where the mean RMSE of BMA-POCRM across simulated trials is 0.0388 lower than for POCRM. In the 5 scenarios where POCRM is a better estimator, the largest difference is of only 0.0075, a five-fold smaller difference. For the average RMSE across all scenarios, shown in Figure \ref{fig:rmse}.D, BMA-POCRM is at  0.0871 whilst POCRM is at just 0.0969. 

As with trials of cohort size 1, Figure \ref{fig:rmse}.B shows that for trials with cohort size of 3 and 6, BMA-POCRM has a lower RMSE for 19 and 20 scenarios respectively. Scenario 13 led to the greatest improvement in RMSE with BMA-POCRM having a 0.0391 and 0.0390 lower RMSE than POCRM for cohort sizes 3 and 6, respectively. The greatest difference in RMSE in favour of POCRM was for scenario 11, where POCRM had a 0.00584 lower RMSE for a cohort size of 3. Similarly, for cohort size of 6, POCRM had best performance relative to BMA-POCRM for scenario 10 where it had an advantage of 0.0058. The negligible differences in RMSE with changes to cohort size is due to the constant total number of patients in each trial, with dose allocations varying depending on cohort size. Hence, leading to some variability in this figure.

\section{Discussion}\label{sec:discussion}

In this paper, the concept of estimation coherency in Phase I clinical trials for drug combinations is introduced. The application of the POCRM and its operating characteristics in this combination setting are explored with the proposition of a novel method, BMA-POCRM. This approach modifies POCRM to take into account uncertainty in dose-toxicity ordering prior to making dose escalations by applying Bayesian model averaging. To evaluate and compare the POCRM and BMA-POCRM, a thorough study of their operating characteristics spanning model accuracy, safety and robustness to estimation incoherency is carried out. The novel method improved on its predecessor in its ability to select correct and acceptable dose levels for progression to Phase II trials. Moreover, the BMA-POCRM also led to fewer overly toxic dose recommendations. The occurrence of estimation incoherencies significantly reduced with the BMA-based method across all scenarios. For example, whilst for POCRM more than 90\% of trials exhibit incoherencies for 21 of 24 scenarios, the worst-case for BMA-POCRM led to only $3.13\%$ of trials with incoherencies. 


The study of operating characteristics shows a clear improvement in performance achieved by BMA-POCRM. Combining estimates of the probability of toxicity across several candidate orderings leads to greater flexibility in potential predictions by the model. POCRM relies on selecting a single most probable ordering on which it bases dose allocations for the current cohort. Thus, BMA-POCRM has a distinct advantage in cases where there is no strong prior knowledge of the underlying dose-toxicity orderings. In particular, the BMA component of the model allows for the specification of intermediary orderings not explicitly included in the original POCRM as shown by the improved RMSE of toxicity estimates across cohort sizes. Despite this, even in scenarios where the correct toxicity ordering is included as a candidate, BMA-POCRM outperforms POCRM. 

Despite the strong performance of the BMA-POCRM in the considered simulations, there are several aspects of this model that have not yet been explored. The novel approach does not completely eliminate the occurrence of estimation incoherencies and the cause of these incoherencies is not clear. Furthermore, a comparison study of the impact of prior information on BMA-POCRM and POCRM performance is also necessary. This is particularly relevant when a correct ordering is included as a candidate, prior information may be available to inform these approaches, potentially leading to favourable performance for POCRM.

\if1\blind
{
\bigskip
    {\noindent \Large\bf Awknowledgements}

This report is independent research supported by the National Institute for Health Research (NIHR Advanced Fellowship, Dr Pavel Mozgunov, NIHR300576). The views expressed in this publication are those of the authors and not necessarily those of the NHS, the National Institute for Health Research or the Department of Health and Social Care (DHSC). T. Jaki and P. Mozgunov received funding from UK Medical Research Council (MC\_UU\_00002\/14 and MC\_UU\_00002\/19, respectively). For the purpose of open access, the author has applied a Creative Commons Attribution (CC BY) licence to any Author Accepted Manuscript version arising.

  \medskip
} \fi

\bibliography{bibliography}

\begin{thebibliography}{}

\bibitem[\protect\citeauthoryear{Barnett, George, Skanji, Saint-Hilary, Jaki, and Mozgunov}{Barnett et~al.}{2024}]{barnett2021comparison}
Barnett, H., M.~George, D.~Skanji, G.~Saint-Hilary, T.~Jaki, and P.~Mozgunov (2024).
\newblock A comparison of model-free phase i dose escalation designs for dual-agent combination therapies.
\newblock {\em Statistical Methods in Medical Research\/}, 09622802231220497.

\bibitem[\protect\citeauthoryear{Cheung}{Cheung}{2019}]{cheung2013package}
Cheung, K. (2019).
\newblock {\em dfcrm: Dose-Finding by the Continual Reassessment Method}.
\newblock R package version 0.2-2.1.

\bibitem[\protect\citeauthoryear{Cheung and Chappell}{Cheung and Chappell}{2002}]{cheung2002simple}
Cheung, Y.~K. and R.~Chappell (2002).
\newblock A simple technique to evaluate model sensitivity in the continual reassessment method.
\newblock {\em Biometrics\/}~{\em 58\/}(3), 671--674.

\bibitem[\protect\citeauthoryear{Gandhi, Bahleda, Tolaney, Kwak, Cleary, Pandya, Hollebecque, Abbas, Ananthakrishnan, Berkenblit, et~al.}{Gandhi et~al.}{2014}]{gandhi2014phase}
Gandhi, L., R.~Bahleda, S.~M. Tolaney, E.~L. Kwak, J.~M. Cleary, S.~S. Pandya, A.~Hollebecque, R.~Abbas, R.~Ananthakrishnan, A.~Berkenblit, et~al. (2014).
\newblock Phase i study of neratinib in combination with temsirolimus in patients with human epidermal growth factor receptor 2--dependent and other solid tumors.
\newblock {\em Journal of Clinical Oncology\/}~{\em 32\/}(2), 68--75.

\bibitem[\protect\citeauthoryear{Ivanova and Kim}{Ivanova and Kim}{2009}]{ivanova2009dose}
Ivanova, A. and S.~H. Kim (2009).
\newblock Dose finding for continuous and ordinal outcomes with a monotone objective function: a unified approach.
\newblock {\em Biometrics\/}~{\em 65\/}(1), 307--315.

\bibitem[\protect\citeauthoryear{Ivanova, Montazer-Haghighi, Mohanty, and D.~Durham}{Ivanova et~al.}{2003}]{ivanova2003improved}
Ivanova, A., A.~Montazer-Haghighi, S.~G. Mohanty, and S.~D.~Durham (2003).
\newblock Improved up-and-down designs for phase i trials.
\newblock {\em Statistics in medicine\/}~{\em 22\/}(1), 69--82.

\bibitem[\protect\citeauthoryear{Ivanova and Wang}{Ivanova and Wang}{2004}]{ivanova2004non}
Ivanova, A. and K.~Wang (2004).
\newblock A non-parametric approach to the design and analysis of two-dimensional dose-finding trials.
\newblock {\em Statistics in Medicine\/}~{\em 23\/}(12), 1861--1870.

\bibitem[\protect\citeauthoryear{Lee and Cheung}{Lee and Cheung}{2009}]{lee2009model}
Lee, S.~M. and Y.~K. Cheung (2009).
\newblock Model calibration in the continual reassessment method.
\newblock {\em Clinical Trials\/}~{\em 6\/}(3), 227--238.

\bibitem[\protect\citeauthoryear{Mozgunov, Gasparini, and Jaki}{Mozgunov et~al.}{2020}]{mozgunov2020surface}
Mozgunov, P., M.~Gasparini, and T.~Jaki (2020).
\newblock A surface-free design for phase i dual-agent combination trials.
\newblock {\em Statistical methods in medical research\/}~{\em 29\/}(10), 3093--3109.

\bibitem[\protect\citeauthoryear{Mozgunov and Jaki}{Mozgunov and Jaki}{2019}]{mozgunov2019information}
Mozgunov, P. and T.~Jaki (2019).
\newblock An information theoretic phase i--ii design for molecularly targeted agents that does not require an assumption of monotonicity.
\newblock {\em Journal of the Royal Statistical Society. Series C, Applied Statistics\/}~{\em 68\/}(2), 347.

\bibitem[\protect\citeauthoryear{Mozgunov, Jaki, Gounaris, Goddemeier, Victor, and Grinberg}{Mozgunov et~al.}{2022}]{mozgunov2022practical}
Mozgunov, P., T.~Jaki, I.~Gounaris, T.~Goddemeier, A.~Victor, and M.~Grinberg (2022).
\newblock Practical implementation of the partial ordering continual reassessment method in a phase i combination-schedule dose-finding trial.
\newblock {\em Statistics in Medicine\/}~{\em 41\/}(30), 5789--5809.

\bibitem[\protect\citeauthoryear{O'Quigley, Pepe, and Fisher}{O'Quigley et~al.}{1990}]{o1990continual}
O'Quigley, J., M.~Pepe, and L.~Fisher (1990).
\newblock Continual reassessment method: a practical design for phase 1 clinical trials in cancer.
\newblock {\em Biometrics\/}, 33--48.

\bibitem[\protect\citeauthoryear{Papoulis and Unnikrishna~Pillai}{Papoulis and Unnikrishna~Pillai}{2002}]{papoulis2002probability}
Papoulis, A. and S.~Unnikrishna~Pillai (2002).
\newblock {\em Probability, random variables and stochastic processes}.

\bibitem[\protect\citeauthoryear{Park and Liu}{Park and Liu}{2020}]{park2020coherence}
Park, Y. and S.~Liu (2020).
\newblock On the coherence of model-based dose-finding designs for drug combination trials.
\newblock {\em Plos one\/}~{\em 15\/}(11), e0242561.

\bibitem[\protect\citeauthoryear{Raftery, Madigan, and Hoeting}{Raftery et~al.}{1997}]{raftery1997bayesian}
Raftery, A.~E., D.~Madigan, and J.~A. Hoeting (1997).
\newblock Bayesian model averaging for linear regression models.
\newblock {\em Journal of the American Statistical Association\/}~{\em 92\/}(437), 179--191.

\bibitem[\protect\citeauthoryear{Riviere, Dubois, and Zohar}{Riviere et~al.}{2015}]{riviere2015competing}
Riviere, M.-K., F.~Dubois, and S.~Zohar (2015).
\newblock Competing designs for drug combination in phase i dose-finding clinical trials.
\newblock {\em Statistics in medicine\/}~{\em 34\/}(1), 1--12.

\bibitem[\protect\citeauthoryear{Wages and Conaway}{Wages and Conaway}{2013}]{wages2013specifications}
Wages, N.~A. and M.~R. Conaway (2013).
\newblock Specifications of a continual reassessment method design for phase i trials of combined drugs.
\newblock {\em Pharmaceutical statistics\/}~{\em 12\/}(4), 217--224.

\bibitem[\protect\citeauthoryear{Wages, Conaway, and O'Quigley}{Wages et~al.}{2011}]{wages2011continual}
Wages, N.~A., M.~R. Conaway, and J.~O'Quigley (2011).
\newblock Continual reassessment method for partial ordering.
\newblock {\em Biometrics\/}~{\em 67\/}(4), 1555--1563.

\bibitem[\protect\citeauthoryear{Yin and Yuan}{Yin and Yuan}{2009}]{yin2009bayesian}
Yin, G. and Y.~Yuan (2009).
\newblock Bayesian dose finding in oncology for drug combinations by copula regression.
\newblock {\em Journal of the Royal Statistical Society: Series C (Applied Statistics)\/}~{\em 58\/}(2), 211--224.

\end{thebibliography}

\newpage

\beginsupplement

\begin{center}
{\large\bf SUPPLEMENTARY MATERIAL}
\end{center}

\normalsize \textbf{ SM.1 \space \space \space SIMULATION SCENARIOS}

\vspace{-0.15em}


\captionsetup{width=\textwidth}

\begin{table}[!ht]
\centering
\begin{adjustbox}{width=\textwidth}
\begin{tabular}{@{}lccccccccccccccc@{}}
\toprule
\multicolumn{2}{c}{} & \multicolumn{14}{c}{\textbf{Drug A}} \\
\multicolumn{2}{c}{\multirow{-2}{*}{\textbf{Dose Level}}} & 1 & 2 & 3 & 4 &  & 1 & 2 & 3 & 4 &  & 1 & 2 & 3 & 4 \\ \midrule
 &  & \multicolumn{4}{c}{\textbf{Scenario 1: 0 CO, 3 OT}} & \textbf{} & \multicolumn{4}{c}{\textbf{Scenario 2: 1 CO, 3 OT}} & \textbf{} & \multicolumn{4}{c}{\textbf{Scenario 3: 2 CO, 3 OT}} \\
 & 1 & 0.01 & 0.04 & 0.16 & 0.16 &  & 0.01 & 0.04 & 0.10 & 0.16 &  & 0.01 & 0.04 & 0.10 & 0.13 \\
 & 2 & 0.07 & 0.13 & 0.20 & {\color[HTML]{3166FF} 0.30} &  & 0.07 & 0.13 & 0.20 & {\color[HTML]{3166FF} 0.30} &  & 0.07 & 0.10 & 0.16 & {\color[HTML]{3166FF} 0.30} \\
 & 3 & 0.10 & 0.20 & {\color[HTML]{3166FF} 0.30} & 0.45 &  & 0.16 & 0.20 & {\color[HTML]{3166FF} 0.30} & 0.40 &  & 0.10 & 0.20 & {\color[HTML]{3166FF} 0.30} & 0.40 \\
 & 4 & 0.20 & {\color[HTML]{3166FF} 0.30} & 0.40 & 0.50 &  & 0.20 & {\color[HTML]{3166FF} 0.30} & 0.45 & 0.50 &  & 0.20 & {\color[HTML]{3166FF} 0.30} & 0.45 & 0.50 \\
 &  & \multicolumn{4}{c}{\textbf{Scenario 4: 0 CO, 6 OT}} & \textbf{} & \multicolumn{4}{c}{\textbf{Scenario 5: 1 CO, 6 OT}} & \textbf{} & \multicolumn{4}{c}{\textbf{Scenario 6: 2 CO, 6 OT}} \\
 & 1 & 0.01 & 0.10 & 0.20 & 0.20 &  & 0.01 & 0.04 & 0.10 & 0.20 &  & 0.01 & 0.05 & 0.15 & 0.20 \\
 & 2 & 0.05 & 0.20 & 0.20 & 0.40 &  & 0.07 & 0.15 & {\color[HTML]{3166FF} 0.30} & 0.40 &  & 0.10 & 0.15 & {\color[HTML]{3166FF} 0.30} & 0.40 \\
 & 3 & 0.15 & {\color[HTML]{3166FF} 0.30} & 0.45 & 0.60 &  & 0.20 & {\color[HTML]{3166FF} 0.30} & 0.45 & 0.55 &  & 0.15 & {\color[HTML]{3166FF} 0.30} & 0.40 & 0.45 \\
 & 4 & {\color[HTML]{3166FF} 0.30} & 0.50 & 0.55 & 0.65 &  & {\color[HTML]{3166FF} 0.30} & 0.50 & 0.60 & 0.65 &  & {\color[HTML]{3166FF} 0.30} & 0.40 & 0.50 & 0.60 \\
 &  & \multicolumn{4}{c}{\textbf{Scenario 7: 0 CO, 10 OT}} &  & \multicolumn{4}{c}{\textbf{Scenario 8: 1 CO, 10 OT}} &  & \multicolumn{4}{c}{\textbf{Scenario 9: 2 CO, 10 OT}} \\
 & 1 & 0.01 & 0.10 & {\color[HTML]{3166FF} 0.30} & 0.40 &  & 0.01 & 0.10 & {\color[HTML]{3166FF} 0.30} & 0.40 &  & 0.10 & 0.05 & {\color[HTML]{3166FF} 0.30} & 0.40 \\
 & 2 & 0.05 & 0.20 & 0.45 & 0.60 &  & 0.05 & 0.20 & 0.45 & 0.60 &  & 0.10 & {\color[HTML]{3166FF} 0.30} & 0.45 & 0.60 \\
 & 3 & 0.15 & 0.50 & 0.65 & 0.80 &  & 0.15 & 0.50 & 0.65 & 0.75 &  & {\color[HTML]{3166FF} 0.30} & 0.50 & 0.60 & 0.65 \\
 & 4 & 0.55 & 0.70 & 0.75 & 0.85 &  & 0.55 & 0.70 & 0.80 & 0.85 &  & 0.55 & 0.60 & 0.70 & 0.75 \\
 &  & \multicolumn{4}{c}{\textbf{Scenario 10: 0 CO, 13 OT}} &  & \multicolumn{4}{c}{\textbf{Scenario 11: 1 CO, 13 OT}} &  & \multicolumn{4}{c}{\textbf{Scenario 12: 2 CO, 13 OT}} \\
 & 1 & 0.15 & {\color[HTML]{3166FF} 0.30} & 0.50 & 0.55 &  & 0.15 & 0.20 & 0.40 & 0.55 &  & 0.15 & 0.20 & 0.40 & 0.45 \\
 & 2 & 0.20 & 0.45 & 0.60 & 0.75 &  & {\color[HTML]{3166FF} 0.30} & 0.45 & 0.60 & 0.75 &  & {\color[HTML]{3166FF} 0.30} & 0.40 & 0.50 & 0.65 \\
 & 3 & 0.40 & 0.65 & 0.80 & 0.96 &  & 0.50 & 0.65 & 0.80 & 0.90 &  & 0.40 & 0.55 & 0.65 & 0.70 \\
 & 4 & 0.70 & 0.85 & 0.90 & 1.00 &  & 0.70 & 0.85 & 0.95 & 1.00 &  & 0.60 & 0.65 & 0.75 & 0.80 \\
 &  & \multicolumn{4}{c}{\textbf{Scenario 1A: 0 CO}} &  & \multicolumn{4}{c}{\textbf{Scenario 2A: 1 CO}} &  & \multicolumn{4}{c}{\textbf{Scenario 3A: 2 CO}} \\
 & 1 & 0.01 & 0.10 & 0.13 & 0.15 &  & 0.01 & 0.05 & 0.10 & 0.18 &  & 0.01 & 0.05 & 0.14 & 0.17 \\
 & 2 & 0.05 & 0.13 & 0.15 & {\color[HTML]{3166FF} 0.30} &  & 0.10 & 0.15 & 0.20 & {\color[HTML]{3166FF} 0.30} &  & 0.10 & 0.14 & 0.20 & {\color[HTML]{3166FF} 0.30} \\
 & 3 & 0.08 & 0.15 & 0.20 & 0.45 &  & 0.18 & 0.20 & 0.40 & 0.44 &  & 0.14 & 0.20 & 0.40 & 0.45 \\
 & 4 & {\color[HTML]{3166FF} 0.30} & 0.40 & 0.50 & 0.55 &  & {\color[HTML]{3166FF} 0.30} & 0.40 & 0.47 & 0.60 &  & {\color[HTML]{3166FF} 0.30} & 0.40 & 0.50 & 0.55 \\
 &  & \multicolumn{4}{c}{\textbf{Scenario 4A: 0 CO}} &  & \multicolumn{4}{c}{\textbf{Scenario 5A: 1 CO}} &  & \multicolumn{4}{c}{\textbf{Scenario 6A: 2 CO}} \\
 & 1 & 0.01 & 0.10 & 0.15 & {\color[HTML]{3166FF} 0.30} &  & 0.05 & 0.15 & {\color[HTML]{3166FF} 0.30} & 0.50 &  & 0.05 & 0.15 & {\color[HTML]{3166FF} 0.30} & 0.45 \\
 & 2 & 0.05 & 0.20 & 0.40 & 0.55 &  & {\color[HTML]{3166FF} 0.30} & 0.40 & 0.55 & 0.75 &  & {\color[HTML]{3166FF} 0.30} & 0.40 & 0.50 & 0.65 \\
 & 3 & {\color[HTML]{3166FF} 0.30} & 0.45 & 0.60 & 0.75 &  & 0.45 & 0.55 & 0.75 & 0.80 &  & 0.40 & 0.55 & 0.65 & 0.70 \\
 & 4 & 0.50 & 0.65 & 0.70 & 0.80 &  & 0.60 & 0.80 & 0.85 & 0.90 &  & 0.60 & 0.65 & 0.75 & 0.80 \\
 &  & \multicolumn{4}{c}{\textbf{Scenario 7A: 0 CO}} &  & \multicolumn{4}{c}{\textbf{Scenario 8A: 1 CO}} &  & \multicolumn{4}{c}{\textbf{Scenario 9A: 2 CO}} \\
 & 1 & 0.05 & {\color[HTML]{3166FF} 0.30} & 0.40 & 0.50 &  & 0.05 & 0.15 & {\color[HTML]{3166FF} 0.30} & 0.45 &  & 0.05 & {\color[HTML]{3166FF} 0.30} & 0.40 & 0.45 \\
 & 2 & 0.15 & 0.40 & 0.55 & 0.65 &  & {\color[HTML]{3166FF} 0.30} & 0.40 & 0.50 & 0.60 &  & 0.15 & 0.40 & 0.50 & 0.60 \\
 & 3 & {\color[HTML]{3166FF} 0.30} & 0.55 & 0.65 & 0.80 &  & 0.50 & 0.50 & 0.65 & 0.70 &  & {\color[HTML]{3166FF} 0.30} & 0.50 & 0.60 & 0.70 \\
 & 4 & 0.60 & 0.70 & 0.75 & 0.85 &  & 0.55 & 0.70 & 0.75 & 0.80 &  & 0.55 & 0.60 & 0.75 & 0.80 \\
 &  & \multicolumn{4}{c}{\textbf{Scenario 13: 15 OT}} &  & \multicolumn{4}{c}{\textbf{Scenario 14}} &  & \multicolumn{4}{c}{\textbf{Scenario 15}} \\
 & 1 & {\color[HTML]{3166FF} 0.30} & 0.40 & 0.50 & 0.57 &  & 0.37 & 0.42 & 0.50 & 0.57 &  & 0.05 & 0.07 & 0.17 & 0.19 \\
 & 2 & 0.40 & 0.48 & 0.57 & 0.68 &  & 0.40 & 0.48 & 0.57 & 0.68 &  & 0.07 & 0.12 & 0.19 & 0.24 \\
 & 3 & 0.45 & 0.61 & 0.72 & 0.87 &  & 0.43 & 0.61 & 0.72 & 0.87 &  & 0.10 & 0.19 & 0.23 & 0.25 \\
\multirow{-40}{*}{\textbf{Drug B}} & 4 & 0.64 & 0.75 & 0.80 & 0.95 &  & 0.64 & 0.75 & 0.85 & 0.95 &  & 0.17 & 0.23 & 0.25 & 0.25 \\ \bottomrule
\end{tabular}
\end{adjustbox}
\caption{scenario settings for all scenarios. Scenarios 1-12 are symmetric, meaning the MTDs are on the same diagonal. Scenarios 1A-9A have asymmetric toxicity matrices. Scenario 14 has all dose levels overly toxic. Scenario 15 has all dose levels below the TTR. Some scenarios have specific design intentions indicated here by the number of correct orderings (CO) and the number of overly toxic doses (OT). The CO number indicates the number of model choices that would have a true ordering for each corresponding setting. The OT number indicates the number of doses that have probability of toxicity greater than $110\%$ of the TTR.}
\label{tab:scen}
\end{table}

\noindent \textbf{\normalsize SM.2 \space \space \space CALIBRATION} \label{sec:calib}


The skeleton is optimised independently for POCRM and BMA-POCRM to ensure a fair comparison. This is done by evaluating performance under scenarios $2$, $6$, $5$A and $13$, which characterise a diverse set of scenarios. Scenario 2 and 6 have 1 and 2 correct orderings, respectively. Scenario 5A has 1 correct ordering and an asymmetric MTD layout. Finally, scenario 13 is where all but one dose level are overly toxic. The geometric mean across simulated trials is considered as it more accurately accounts for poor performance relative to the arithmetic mean. For each scenario, $5,000$ trials are simulated for which the results are aggregated by taking the geometric mean of the PCS across all $20,000$ trials. Trial simulations have a total of $60$ patients with $60$ cohorts of size $1$. The skeleton with the highest geometric mean PCS and lowest standard deviation of PCS are selected for the complete simulation study. 

Two approaches for generating the underlying probability skeleton are considered. First, the indifference interval protocol defined in \cite{lee2009model} and implemented by the \verb|dfcrm| \verb|R| package \citep{cheung2013package}. This approach has two parameters, the prior MTD and the $\delta$ parameter, which controls the distance between the prior probability of toxicity for adjacent doses in the skeleton. Grid optimisation across these parameters is carried out to select the optimal indifference interval skeleton.

A linear skeleton is also implemented, whereby the prior probability of toxicity at the lowest dose-level is given, and the skeleton increases at a fixed interval. Recursively, this is defined as, $$\widehat{R}^{(0)}(d_i) = 
\begin{cases}
         p_0 & \text{if $i=1$,}\\
         \widehat{R}^{(0)} (d_{i-1}) + \eta & \text{otherwise,}
    \end{cases} $$ where $\widehat{R}^{(0)}(d_i)$
 is the prior probability of toxicity for dose $i$, $p_0$ is the prior probability of toxicity for the lowest dose, or \emph{start probability} and $\eta$ is the fixed increment, or \emph{spacing}. Again, grid optimisation is carried out to find the best linear skeleton.

The calibration results for POCRM under an indifference interval are presented in Figure S1. These results show that for many prior MTDs, selecting delta as 0.03 leads to a misspecified skeleton. The same trend was empirically observed for greater values of delta. In particular, when delta is set to 0.04 a diverging marginal likelihood interval is observed for all settings of the prior MTD in at least one scenario. Since the indifference interval protocol begins at the prior MTD and obtains each adjacent skeleton value sequentially, if $\delta$ is too high, the protocol will reach the bounds of 0 and 1 quickly, resulting in several adjacent values which are arbitrarily close to 0 and 1. For example, selecting $d_{15}$ as the prior MTD and $\delta = 0.04$ the following skeleton is obtained, $$\{1.94e^{-12}, 4.17e^{-10}, 3.08e^{-08}, \ldots, 0.22, 0.30, 0.38\}.$$ A misspecified skeleton leads to a diverging marginal likelihood integral and is observed both for BMA-POCRM and POCRM as they share a common approach to computing the marginal likelihood in Equation (\ref{eq:marginallikelihood}). The skeletons proposed at this level are frequently unintuitive and result in small differences between dose levels at the boundaries near 0 and 1. As a result, $\delta \geq 0.03$ is excluded for the indifference interval skeleton and $\eta \; \text{(spacing)} \; = 0.04, 0.05, 0.06$ for the linear skeleton due to the missing values shown in Figures S1.A and S1.C, respectively.  

Under the indifference interval skeleton, POCRM exhibits greatest PCS with a prior MTD \(d_1\) and $\delta = 0.02$. Switching to a prior MTD \(d_2\) leads to a slightly lower PCS but also lower standard deviation in performance. This is favourable as it implies greater consistency across scenarios. Assuming a linear skeleton, the design performs optimally under $p_0=0.4$ and $\eta = 0.03$, however, this presents significantly greater SD than when $p_0=0.01$ and $\eta = 0.03$. Overall, the linear skeleton is much less consistent across parametrisations when compared to the indifference interval method for the POCRM.

Similarly, Figure S2 shows that under the indifference interval for BMA-POCRM, optimal performance is observed according to mean PCS at a prior MTD \(d_2\) and $\delta=0.02$. This setting also leads to the smallest SD across all valid parametrisations of the indifference interval (i.e. excluding those where $\delta > 0.03$).  This setting of the indifference interval also outperforms all given parametrisations of the linear skeleton according to mean PCS. The linear skeleton with $\eta = 0.03$ and $p_0 = 0.01$ leads to lower SD, however, this comes at the expense of a point reduction in PCS. 

Moving forward, a matching skeleton is selected for both POCRM and BMA-POCRM. The indifference interval skeleton with prior MTD  \(d_2\) and $\delta = 0.02$ is used due to the balance between aggregate performance and consistency across both the methods included (i.e. comparatively low SD). 

\begin{figure}
    \centering
    \includegraphics[width=0.75\textwidth]{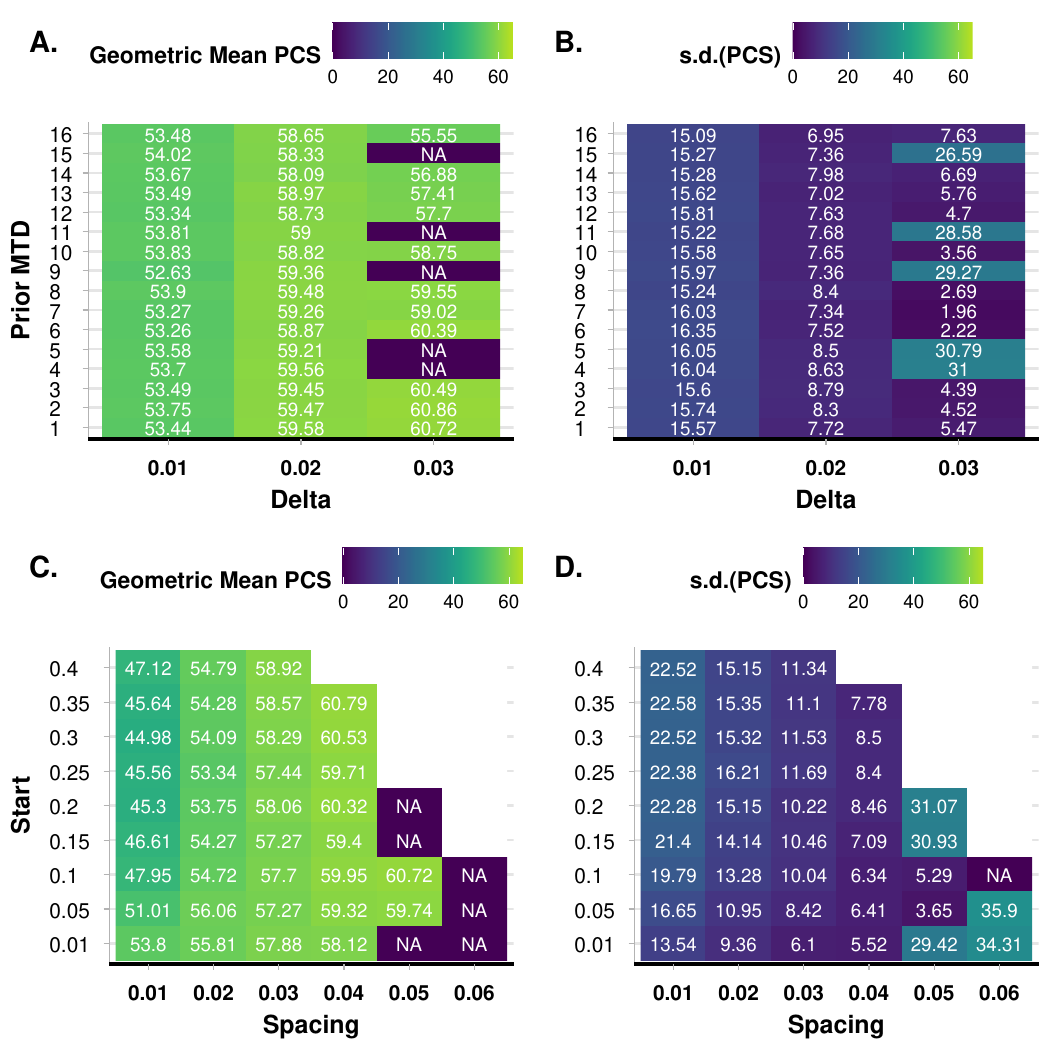}
    \caption{Results for prior calibration of skeleton parameters for the POCRM with model selection. Panel A. Geometric mean of PCS for various values of the indifference interval skeleton under the POCRM model. Panel B. Standard deviation of PCS across scenarios selected for calibration when applying the indifference interval skeleton under the POCRM model. Panel C. Geometric mean of PCS for various values of the linear skeleton under the POCRM model. Panel D. Standard deviation of PCS across scenarios selected for calibration when applying the linear skeleton under the POCRM model. Missing values indicate at least one diverging marginal likelihood integral over simulations.}
    \label{fig:pocrm_calib}
\end{figure}

\begin{figure}
    \centering
    \includegraphics[width=0.75\textwidth]{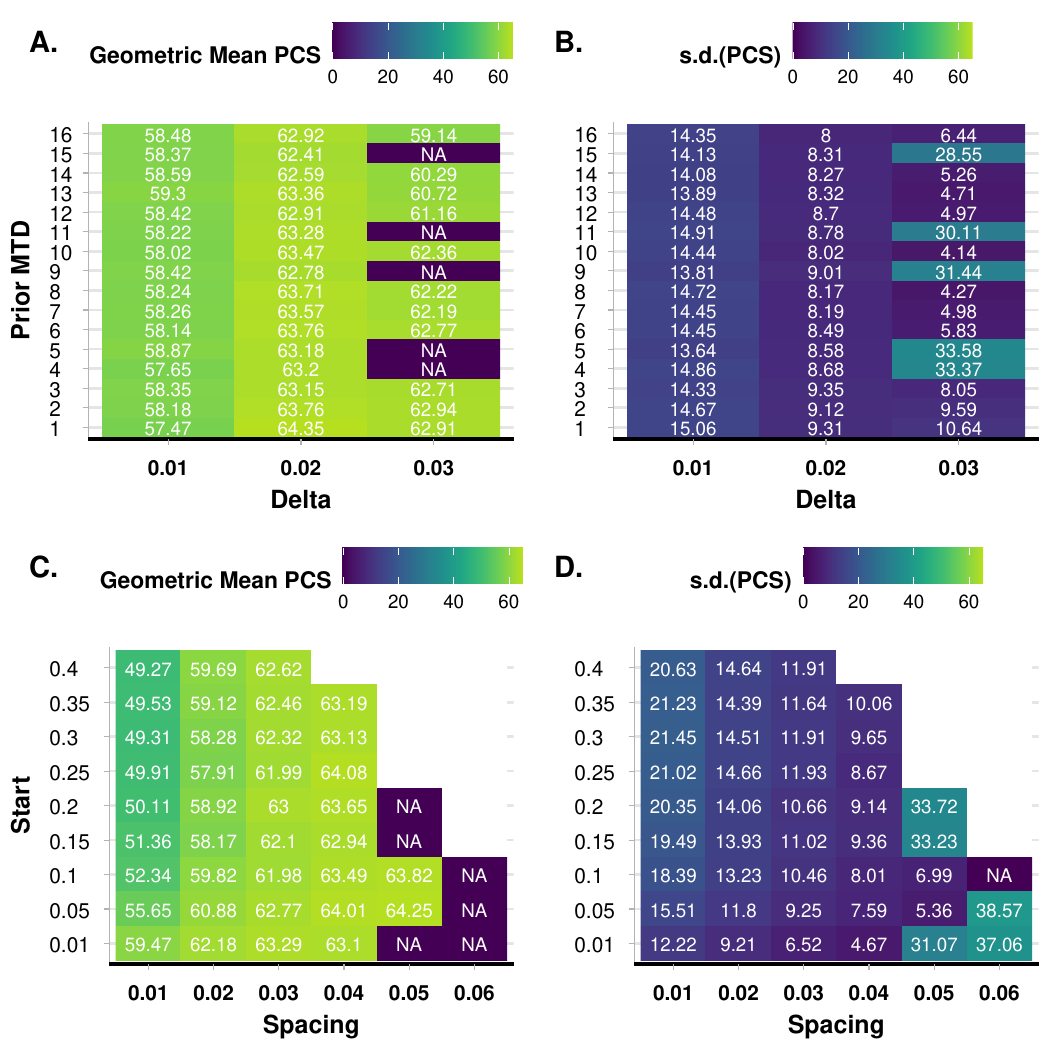}
    \caption{Results for prior calibration of skeleton parameters for the BMA-POCRM. Panel A. Geometric mean of PCS for various values of the indifference interval skeleton under the BMA-POCRM model. Panel B. Standard deviation of PCS across scenarios selected for calibration when applying the indifference interval skeleton under the BMA-POCRM model. Panel C. Geometric mean of PCS for various values of the linear skeleton under the BMA-POCRM model. Panel D. Standard deviation of PCS across scenarios selected for calibration when applying the linear skeleton under the BMA-POCRM model. Missing values indicate at least one diverging marginal likelihood integral over simulations.}
    \label{fig:bma_calib}
\end{figure}

\bigskip

\newpage

\noindent \normalsize \textbf{SM.3 \space \space \space VARIANCE CALIBRATION}\label{apx:calib}

The effect of altering the variance of the prior distribution of the $a$ parameter of the working model is also explored in Supplementary Materials. It was found that the prior variance does not have any influence on the mean PCS for neither of the two approaches, and hence the value of 1.34 is used for further simulations.

\newpage

\noindent \normalsize \textbf{SM.4 \space \space \space FURTHER SIMULATION RESULTS}

\begin{figure}[h!]
    \centering
    \includegraphics[width=0.75\textwidth]{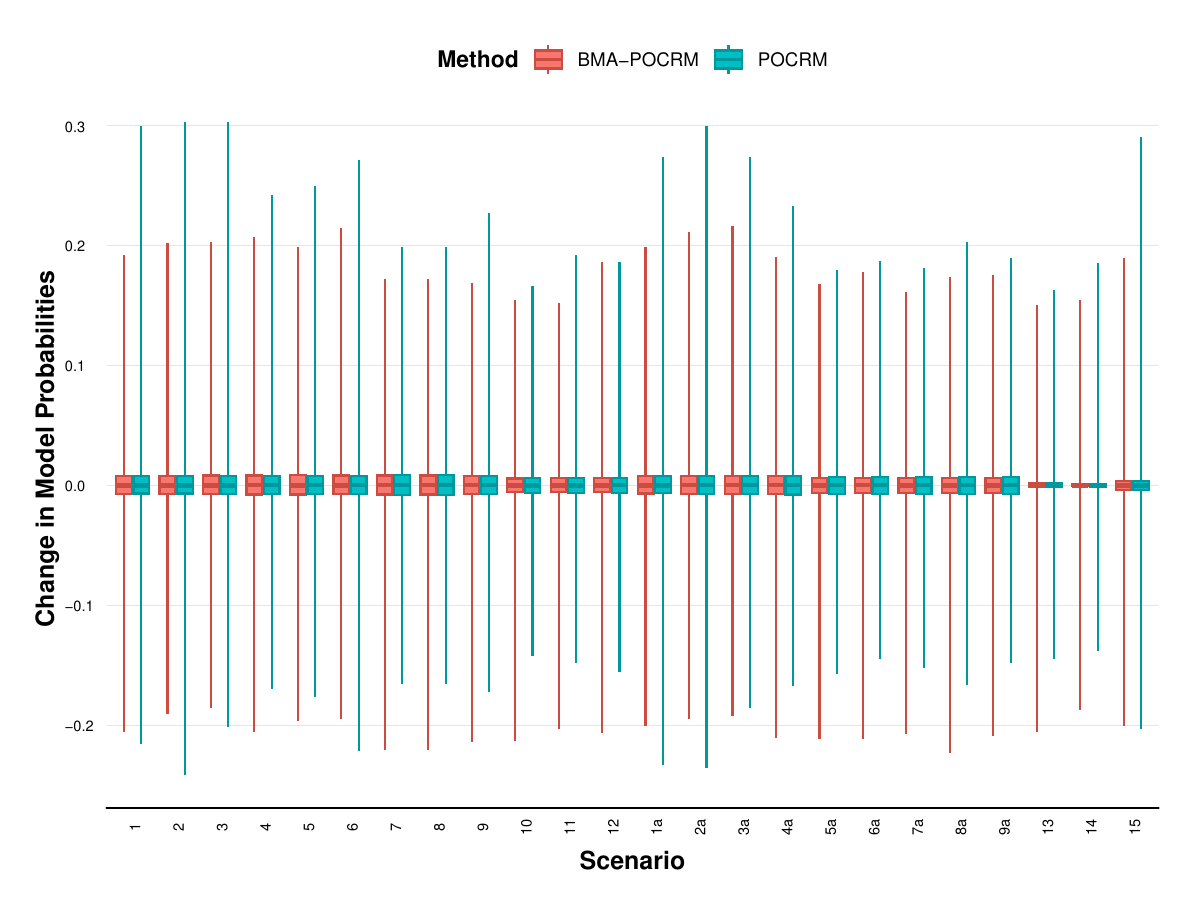}
    \caption{Distribution of changes in posterior model probabilities following the induction of each cohort for simulated trials with cohort size 1. }
    \label{fig:tox_change_coh1}
\end{figure}

\noindent \normalsize \textbf{SM.5 \space \space \space TRIAL DATA}\label{apx:realtrial}

\begin{table}[h!]
\centering
\caption{Real trial data generated by \cite{gandhi2014phase} where each cell represents $y_j/n_j$ the number of observed DLTs and the number of patients assigned to each combination.}
\label{tab:realtrialsim_dlts}
\begin{tabular}{@{}llllll@{}}
\toprule
               & \multicolumn{5}{l}{Temsirolimus (mg)}                                \\
               &              & \textit{15} & \textit{25} & \textit{50} & \textit{75} \\ \midrule
               & \textit{120} & 0/2         & 0/4         & 1/5         & 0/4         \\
Neratinib (mg) & \textit{160} & 1/4         & 1/4         & 0/5         & 3/6         \\
               & \textit{200} & 0/4         & 1/8         & 1/2         & --          \\
               & \textit{240} & 2/4         & --          & --          & --          \\ \bottomrule
\end{tabular}%
\end{table}

\begin{table}[h!]
\centering
\caption{Key for doses allocated in the \cite{gandhi2014phase} trial.}
\label{tab:realtrialsim_alloc}
\begin{tabular}{@{}llllll@{}}
\toprule
               & \multicolumn{5}{l}{Temsirolimus (mg)}                                \\
               &              & \textit{15} & \textit{25} & \textit{50} & \textit{75} \\ \midrule
               & \textit{120} & $d_{1}$         & $d_{2}$         & $d_{3}$         & $d_{4}$         \\
Neratinib (mg) & \textit{160} & $d_{5}$         & $d_{6}$         & $d_{7}$         & $d_{8}$         \\
               & \textit{200} & $d_{9}$         & $d_{10}$         & $d_{11}$         & --          \\
               & \textit{240} & $d_{12}$         & --          & --          & --          \\ \bottomrule
\end{tabular}%
\end{table}

\begin{table}[]
\centering
\caption{Dose allocations with POCRM in the motivating trial \citep{gandhi2014phase} where each cell represents $y_j/n_j$ the number of observed DLTs and the number of patients assigned to each combination. The starting combination is $d_5$.}
\label{tab:realtrialsim_alloc_pocrm}
\begin{tabular}{@{}cccccc@{}}
\toprule
                                &              & \multicolumn{4}{c}{Temsirolimus (mg)}                 \\
                                &              & \textit{15} & \textit{25} & \textit{50} & \textit{75} \\ \midrule
\multirow{4}{*}{Neratinib (mg)} & \textit{120} & 0/0         & 0/0         & 0/1         & 0/2         \\
                                & \textit{160} & 0/1         & 0/0         & 0/1         & 1/1         \\
                                & \textit{200} & 0/1         & 1/16        & 0/0         & --          \\
                                & \textit{240} & 15/30       & --          & --          & --          \\ \midrule 
\end{tabular}%
\end{table}

\begin{table}[]
\centering
\caption{Dose allocations with BMA-POCRM in the motivating trial \citep{gandhi2014phase} where each cell represents $y_j/n_j$ the number of observed DLTs and the number of patients assigned to each combination. The starting combination is $d_5$.}
\label{tab:realtrialsim_alloc_bma}
\begin{tabular}{@{}cccccc@{}}
\toprule
                                &              & \multicolumn{4}{c}{Temsirolimus (mg)}                 \\
                                &              & \textit{15} & \textit{25} & \textit{50} & \textit{75} \\ \midrule
\multirow{4}{*}{Neratinib (mg)} & \textit{120} & 0/0         & 0/0         & 0/0         & 0/2         \\
                                & \textit{160} & 0/1         & 0/0         & 0/0         & 1/1         \\
                                & \textit{200} & 0/1         & 1/17        & 1/2         & --          \\
                                & \textit{240} & 13/28       & --          & --          & --          \\ \midrule 
\end{tabular}%
\end{table}

\end{document}